\documentclass[prb,twocolumn,showpacs,amsmath,amssymb]{revtex4}

\usepackage{dcolumn}
\usepackage{bm}
\usepackage{epsfig}

\begin{document}


\title{Spin injection and spin accumulation in all-metal mesoscopic spin valves\\}

\author{F.J. Jedema}%
\email{jedema@phys.rug.nl} %
\author{M.S.Nijboer}%
\author{A.T. Filip}%
\altaffiliation{Present address: Department of Applied Physics and Center for
Nanomaterials, Eindhoven University of Technology, 5600 MB, The Netherlands.}
\author{B.J. van Wees}%
\affiliation{Department of Applied Physics and Materials Science Center,
University of Groningen,\\ Nijenborgh 4, 9747 AG Groningen, The Netherlands}
\date{\today}

\begin{abstract}

We study the electrical injection and detection of spin accumulation in lateral
ferromagnetic metal-nonmagnetic metal-ferromagnetic metal (F/N/F) spin valve
devices with transparent interfaces. Different ferromagnetic metals, permalloy
(Py), cobalt (Co) and nickel (Ni), are used as electrical spin injectors and
detectors. For the nonmagnetic metal both aluminium (Al) and copper (Cu) are
used. Our multi-terminal geometry allows us to experimentally separate the spin
valve effect from other magneto resistance signals such as the anomalous
magneto resistance (AMR) and Hall effects. We find that the AMR contribution of
the ferromagnetic contacts can dominate the amplitude of the spin valve effect,
making it impossible to observe the spin valve effect in a 'conventional'
measurement geometry. In a 'non local' spin valve measurement we are able to
completely isolate the spin valve signal and observe clear spin accumulation
signals at $T=~4.2$ K as well as at room temperature (RT). For aluminum we
obtain spin relaxation lengths ($\lambda_{sf}$) of $1.2$ $\mu$m and $600$ nm at
$T=4.2$ K and RT respectively, whereas for copper we obtain $1.0$ $\mu$m and
$350$ nm. The spin relaxation times $\tau_{sf}$ in Al and Cu are compared with
theory and results obtained from giant magneto resistance (GMR), conduction
electron spin resonance (CESR), anti-weak localization and superconducting
tunneling experiments. The spin valve signals generated by the Py electrodes
($\alpha_F\lambda_F=0.5~[1.2]$ nm at RT [$T=4.2$ K]) are larger than the Co
electrodes ($\alpha_F\lambda_F=0.3~[0.7]$ nm at RT [$T=4.2$ K]), whereas for Ni
($\alpha_F\lambda_F<0.3$ nm at RT and $T=4.2$ K) no spin signal is observed.
These values are compared to the results obtained from GMR experiments.
\end{abstract}

\maketitle

\section{introduction}
Spintronics is a rapidly emerging field in which one tries to study or make
explicit use of the spin degree of freedom of the
electron.\cite{spinbook,prinz1,wolf1} Sofar, the most well known examples of
spintronics are the giant magneto resistance (GMR) of metallic
multi-layers\cite{gijs1,ans1,bass1} and tunneling magneto resistance (TMR) of
magnetic tunnel junctions\cite{tedrow1,moodera1}. Injection of hot electrons
$\approx 1~eV$ above the Fermi energy $E_F$ in Co/Cu (multi) layers have shown
a significant spin filtering effect, enabling transistor functionality and
ballistic electron magnetic microscopy.\cite{monsma1,rippard1} Recent
experiments have shown the ability of spin polarized currents to initiate a
(local) magnetization reversal in thin ferromagnetic wires and Co/Cu
multi-layer pillars\cite{slons1,sun1,wegrowe1,myers1,grollier1}. A new
direction is emerging, where one actually wants to inject spin currents,
transfer and manipulate the spin information at the Fermi-level, and detect the
resulting spin polarization in nonmagnetic metals and semiconductors. Because
of spin-orbit interaction, the electron spin can be flipped and consequently a
spin polarized current will have a finite lifetime. For this reason it is
necessary to study spin transport in systems, where the 'time of flight' of the
electrons between the injector and detector is shorter than the spin relaxation
time. A first and successful attempt to electrically inject and detect spins in
metals dates back to 1985 when Johnson and Silsbee successfully demonstrated
spin accumulation in a single crystal aluminium bar up to temperatures of 77 K.
\cite{silsbee1,silsbee2} In their pioneering experiments they were able to
observe spin precession of the induced non-equilibrium magnetization, made
possible by the long spin relaxation lengths $\lambda_{sf}
>50~\mu$m. In (diffusive) thin metallic films however, the spin relaxation
length corresponds to typical length scales of 1 $\mu$m. We use a lateral
mesosopic spin valve, to access and probe this length
scale\cite{jedema1,jedema2,jedema3,jedema4}. We note that a similar experiment
using planar spin valves has been reported in Ref. \onlinecite{birge1}.

In Sec. \ref{section2} a review of the basic model for spin transport in the
diffusive transport regime is given, whereas in Sec. \ref{section3} this model
is applied to our multi-terminal device geometry. A multi-terminal resistor
model of spin injection and detection is presented in Sec. \ref{section4} in
order to elucidate the principles behind the reduction of the polarization of
the spin current at a transparent F/N interface, also referred to as
"conductivity mismatch".\cite{schmidt} The sample fabrication process and
measurement geometry are described in Sec. \ref{geometry}. Spin accumulation
measurements in a 'conventional' and 'non-local' geometry for Py/Cu/Py and
Py/Al/Py spin valves will be presented in Sec. \ref{pycu} and Sec. \ref{pyal},
whereas spin accumulation measurements on Co/Cu/Co and Ni/Cu/Ni spin valves
will be presented in Sec. \ref{CoNi}. In Sec. \ref{spinrelaxation} the obtained
results of Secs. \ref{pycu},\ref{pyal} and \ref{CoNi} are analyzed using the
model for spin transport in the diffusive regime and the results are compared
to GMR, CESR, anti-weak localization and superconducting tunneling experiments.

\section{Theory of spin injection and accumulation} \label{section2}

In general, electron transport through a diffusive channel is a result of a
difference in the (electro-)chemical potential of two connected electron
reservoirs.\cite{datta} An electron reservoir is an electron bath in full
thermal equilibrium. The chemical potential $\mu_{ch}$ is by definition the
energy needed to add one electron to the system, usually set to zero at the
Fermi energy (this convention is adapted throughout this text), and accounts
for the kinetic energy of the electrons. In the linear response regime, i.e.
for small deviations from equilibrium ($|eV| < kT$), the chemical potential
equals the excess electron density $n$ divided by the density of states at the
Fermi energy, $\mu_{ch} =n/N(E_F)$. In addition an electron may also have an
potential energy, e.g. due to the presence of an electric field $\textbf{E}$.
The additional potential energy for a reservoir at potential $V$ should be
added to $\mu_{ch}$ in order to obtain the electrochemical potential (in the
absence of a magnetic field):

\begin{equation}
\mu=\mu_{ch}-eV \; , \label{ChemPot}
\end{equation}
where $e$ denotes the absolute value of the electron charge.

From eq. [\ref{ChemPot}] it is clear that a gradient of $\mu$, the driving
force of electron transport, can result from either a spatial varying electron
density $\nabla n$ or an electric field $\textbf{E}=-\nabla V$. Since $\mu$
fully characterizes the reservoir one is free to describe transport either in
terms of diffusion ($\textbf{E}=0$, $\nabla n\neq 0$) or in terms of electron
drift ($\textbf{E}\neq 0$, $\nabla n= 0$). In the drift picture the whole Fermi
sea has to be taken into account and consequently one has to maintain a
constant electron density everywhere by imposing: $\nabla n= 0$. We use the
diffusive picture where only the energy range $\Delta \mu$, the difference in
the electrochemical potential between the two reservoirs, is important to
describe transport. Both approaches (drift and diffusion) are equivalent in the
linear regime and are related to each other via the Einstein relation:

\begin{equation}
\sigma=e^2N(E_F)D \; , \label{einstein}
\end{equation}
where $\sigma$ is the conductivity and $D$ the diffusion constant.

We focus on the diffusive transport regime, which applies when the mean free
path $l_e$ is shorter than the device dimensions. The description of electrical
transport in a ferromagnet in terms of a two-current (spin-up and spin-down)
model dates back to Fert and Campbell \cite{fert1}. Van Son \emph{et al.}
\cite{son1} have extended the model to describe transport through
ferromagnet-nonmagnetic metal interfaces. A firm theoretical underpinning,
based on the Boltzmann transport equation has been given by Valet and
Fert.\cite{valet1} They have applied the model to describe the effects of spin
accumulation and spin dependent scattering on the GMR effect in magnetic
multilayers. This standard model allows for a detailed quantitative analysis of
the experimental results.

An alternative model, based on thermodynamic considerations, has been put
forward and applied by Johnson and Silsbee (JS)\cite{silsbee3}. In principle
both models describe the same physics, and should therefore be equivalent.
However, the JS model has a drawback in that it does not allow a direct
calculation of the spin polarization of the current ($\eta$ in
Refs.\onlinecite{silsbee1,silsbee3,john1,john2}), whereas in the standard model
all measurable quantities can be directly related to the parameters of the
experimental system.\cite{valet1,fert2,herschfield}

The transport in a ferromagnet is described by spin dependent conductivities:

\begin{eqnarray}
\sigma_\uparrow & = & N_\uparrow e^2 D_\uparrow, \;\textrm{with } D_\uparrow =
\frac{1}{3} v_{F\uparrow} l_{e\uparrow}\
\label{conductivityup}\\
\sigma_\downarrow & = & N_\downarrow e^2 D_\downarrow,\;\textrm{with }
D_\downarrow = \frac{1}{3} v_{F\downarrow} l_{e\downarrow}\;,
\label{conductivitydown}
\end{eqnarray}

where $N_{\uparrow,\downarrow}$ denotes the spin dependent density of states
(DOS) at the Fermi energy ($E_F$), and $D_{\uparrow,\downarrow}$ the spin
dependent diffusion constants, expressed in the spin dependent Fermi velocities
$v_{F\uparrow,\downarrow}$, and electron mean free paths $l_{e
\uparrow,\downarrow}$. Throughout this paper our notation is $\uparrow$ for the
majority spin direction and $\downarrow$ for the minority spin direction. Note
that the spin dependence of the conductivities is determined by \emph{both}
density of states and diffusion constants. This should be contrasted with
magnetic F/I/F or F/I/N tunnel junctions, where the spin polarization of the
tunneling electrons is determined by the spin-dependent (local)
DOS.\cite{tedrow1,stearns1,oleinik1} Also in a typical ferromagnet several
bands (which generally have different spin dependent density of states and
effective masses) contribute to the transport. However, provided that the
elastic scattering time and the interband scattering times are shorter than the
spin flip times (which is usually the case) the transport can still be
described in terms of well defined spin up and spin down conductivities.

Because the spin up and spin down conductivities are different, the current in
the bulk ferromagnet will be distributed accordingly over the two spin
channels:

\begin{eqnarray}
j_\uparrow & = & \frac{\sigma_{\uparrow}}{e}\frac
{\partial\mu_{\uparrow}}{\partial x}
\label{currentup}\\
j_\downarrow & = & \frac{\sigma_{\downarrow}}{e}\frac
{\partial\mu_{\downarrow}}{\partial x}\;, \label{currentdown}
\end{eqnarray}

where $j_{\uparrow\downarrow}$ are the spin up and spin down current densities.
According to Eqs. \ref{currentup} and \ref{currentdown}  the current flowing in
a bulk ferromagnet is spin polarized, with a polarization given by:

\begin{equation}
\alpha_F=\frac{\sigma_\uparrow-\sigma_\downarrow}{\sigma_\uparrow+\sigma_\downarrow}.
\label{polarization}
\end{equation}

The next step is the introduction of spin flip processes, described by a spin
flip time $\tau_{\uparrow\downarrow}$ for the average time to flip an up-spin
to a down-spin, and $\tau_{\downarrow\uparrow}$ for the reverse process. The
detailed balance principle imposes that
$N_\uparrow/\tau_{\uparrow\downarrow}=N_\downarrow/\tau_{\downarrow\uparrow}$,
so that in equilibrium no net spin scattering takes place. As pointed out
already, usually these spin flip times are larger than the momentum scattering
time $\tau_e=l_e/v_F$. The transport can then be described in terms of the
parallel diffusion of the two spin species, where the densities are controlled
by spin flip processes. It should be noted however that in particular in
ferromagnets (e.g. permalloy\cite{dub1,steenwyk1,holody1}) the spin flip times
may become comparable to the momentum scattering time. In this case an
(additional) spin-mixing resistance arises \cite{fert3,fert4,gijs1}, which we
will not discuss further here.

The effect of the spin flip processes can now be described by the following
equation (assuming diffusion in one dimension only):
\begin{equation}
D\frac{\partial^2 (\mu_\uparrow-\mu_\downarrow)}{\partial
x^2}=\frac{(\mu_\uparrow-\mu_\downarrow)}{\tau_{sf}}, \label{diffusion}
\end{equation}

where $D=D_\uparrow D_\downarrow (N_\uparrow+N_\downarrow) /(N_\uparrow
D_\uparrow + N_\downarrow D_\downarrow)$ is the spin averaged diffusion
constant, and the spin relaxation time $\tau_{sf}$ is given by: $1/\tau_{sf}=
1/\tau_{\uparrow\downarrow} + 1/\tau_{\downarrow\uparrow}$. We note that
$\tau_{sf}$ represents the timescale over which the non-equilibrium spin
accumulation ($\mu_\uparrow-\mu_\downarrow$) decays and therefore is equal to
the spin lattice relaxation time $T_1$ used in the Bloch equations:
$\tau_{sf}=T_1$ .\cite{notation1} Using the requirement of current
conservation, the general solution of eq. \ref{diffusion} for a uniform
ferromagnet or nonmagnetic wire is now given by:

\begin{eqnarray}
\mu_\uparrow & = & A+Bx+\frac{C}{\sigma_\uparrow}
exp(-x/\lambda_{sf})+\frac{D}{\sigma_\uparrow} exp(x/\lambda_{sf})
\label{solutionup} \\
\mu_\downarrow & = & A+Bx-\frac{C}{\sigma_\downarrow}
exp(-x/\lambda_{sf})-\frac{D}{\sigma_\downarrow} exp(x/\lambda_{sf})\;,
\label{solutiondown}
\end{eqnarray}

where we have introduced the spin relaxation length
$\lambda_{sf}=\sqrt{D\tau_{sf}}$. The coefficients A,B,C, and D are determined
by the boundary conditions imposed at the junctions where the wires are coupled
to other wires. In the absence of an interface resistance and spin flip
scattering at the interfaces, the boundary conditions are: 1) continuity of
$\mu_\uparrow$, $\mu_\downarrow$ at the interface, and 2) conservation of
spin-up and spin-down currents $j_\uparrow$, $j_\downarrow$ across the
interface.

\section{Spin accumulation in multi-terminal spin valve
structures} \label{section3}

\begin{figure}[htb]
\centerline{\psfig{figure=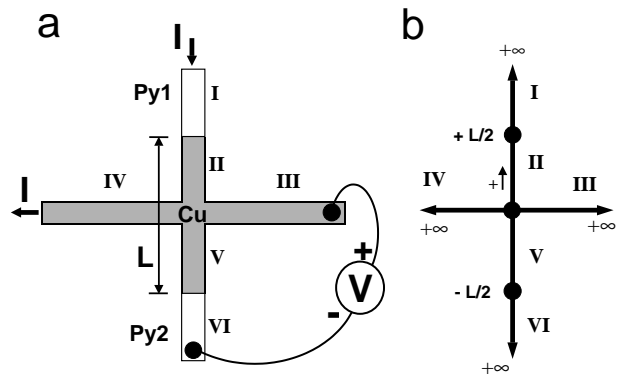,width=86mm}} \caption{(a) Schematic
representation of the multi-terminal spin valve device. Regions \emph{I} and
\emph{VI} denote the injecting ($F_1$) and detecting ($F_2$) ferromagnetic
contacts, whereas regions \emph{II} to \emph{V} denote the four arms of a
normal metal cross ($N$) placed in between the two ferromagnets. A spin
polarized current is injected from region \emph{I} into region \emph{II} and
extracted at region \emph{IV}. (b) Diagram of the electrochemical potential
solutions (Eqs. \ref{solutionup} and \ref {solutiondown}) in each of the six
regions of the multi-terminal spin valve. The nodes represent the origins of
the coordinate axis in the 6 regions, the arrows indicate the (chosen)
direction of the positive x-coordinate. Regions \emph{II} and \emph{III} have a
finite length of half the Py electrode spacing $L$. The other regions are
semi-infinite.} \label{diagram}
\end{figure}

We will now apply the model of spin injection to a non local geometry, which
reflects our measurement and device geometry, see Fig.\ref{diagram}a and
Fig.\ref{sample}c.

In our (1-dimensional) geometry we can identify 6 different regions for which
Eqs. \ref{solutionup} and \ref{solutiondown} have to be solved according to
their boundary conditions at the interface. The geometry is schematically shown
in Fig.\ref{diagram}b, where the 6 different regions are marked with roman
letters I to VI. According to Eq. \ref{solutionup} the equations for the
spin-up electrochemical potentials in these regions, assuming parallel
magnetization of the ferromagnetic regions, read:

\begin{align*}
\mu_\uparrow
&=~A-\frac{je}{\sigma_F}x+\frac{2C}{\sigma_{F}(1+\alpha_{F})}exp(-x/\lambda_{F})
\tag{$I$}
\\
\mu_\uparrow
&=~\frac{-je}{\sigma_N}x+\frac{2E}{\sigma_{N}}exp(-x/\lambda_{N})+\frac{2F}{\sigma_{N}}
exp(x/\lambda_{N}) \tag{$II$}
\\
\mu_\uparrow &=~\frac{2G}{\sigma_{N}}exp(-x/\lambda_{N}) \tag{$III$}
\\
\mu_\uparrow &=~\frac{je}{\sigma_N}x+\frac{2G}{\sigma_{N}}exp(-x/\lambda_{N})
\tag{$IV$}
\\
\mu_\uparrow &=~\frac{2H}{\sigma_{N}}exp(-x/\lambda_{N})+\frac{2K}{\sigma_{N}}
exp(x/\lambda_{N}) \tag{$V$}
\\
\mu_\uparrow &=~B+\frac{2D}{\sigma_{F}(1+\alpha_{F})}exp(-x/\lambda_{F})\;,
\tag{$VI$}
\end{align*}

where we have written $\sigma_\uparrow=\sigma_F(1+\alpha_F)/2$ and
$A,B,C,D,E,F,G,H$ and $K$ are 9 unknown constants. The equations for the
spin-down electrochemical potential in the six regions of fig. \ref{diagram}
can be found by putting a minus sign in front of the constants $C,D,E,F,H,K,G$
and $\alpha_F$ in Eqs. $I$ to $VI$. Constant $B$ is the most valuable to
extract from this set of equations, for it gives directly the difference
between the electrochemical potential measured with a normal metal probe at the
center of the nonmagnetic metal cross in fig.\ref{diagram}a and the
electrochemical potential measured with a ferromagnetic voltage probe at the
F/N interface of region $V$ and $VI$. For $\lambda_{sf}>>L$ i.e. no spin
relaxation in the nonmagnetic metal of regions II and V, the ferromagnetic
voltage probe effectively probes the electrochemical potential difference
between spin-up and spin-down electrons at center of the nonmagnetic metal
cross. Solving the Eqs. $I$ to $VI$ by taking the continuity of the spin-up and
spin-down electrochemical potentials and the conservation of spin-up and spin
down-currents at the 3 nodes of Fig. \ref{diagram}b, one obtains:

\begin{equation}
B=-je\frac{\alpha_F^2\frac{\lambda_N}{\sigma_N}e^{-L/2\lambda_N}}
{2(M+1)[Msinh(L/2\lambda_N)+cosh(L/2\lambda_N)]}\;, \label{constantb}
\end{equation}
where $M=(\sigma_F\lambda_N/\sigma_N\lambda_F)(1-\alpha_F^2)$ and $L$ is the
length of the nonmagnetic metal strip in between the ferromagnetic electrodes.
The magnitude of the spin accumulation at the F/N interface of region $V$ and
$VI$ is given by: $\mu_\uparrow - \mu_\downarrow=B/\alpha_F$.

In the situation where the ferromagnets have an anti-parallel magnetization
alignment, the constant $B$ of Eq. \ref{constantb} gets a minus sign in front .
Upon changing from parallel to anti-parallel magnetization configuration (a
spin valve measurement) a difference of $\Delta\mu=~2B$ will be detected in
electrochemical potential between the normal metal and ferromagnetic voltage
probe. This leads to the definition of the so-called spin-coupled or
spin-dependent resistance $\Delta R=~\frac{2B}{-ejS}$, where $S$ is the
cross-sectional area of the nonmagnetic strip:

\begin{equation}
\Delta R = \frac{\alpha_F^2\frac{\lambda_N}{\sigma_NS}e^{-L/2\lambda_N}}{(M+1)
[Msinh(L/2\lambda_N)+cosh(L/2\lambda_N)]}\;. \label{Rspinfull}
\end{equation}

Equation \ref{Rspinfull} shows that for $\lambda_N << L$, the magnitude of the
spin signal $\Delta R$ will decay exponentially as a function of L. In the
opposite limit, $\lambda_F << L << \lambda_N$ the spin signal $\Delta R$ has a
1/L dependence. In this limit and under the constraint that $ML/2\lambda_N>>1$,
we can write Eq. \ref{Rspinfull} as:

\begin{equation}
\Delta R = \frac{2\alpha_F^2\lambda_N^2}{M(M+1)\sigma_NSL}\;.
 \label{Rspin}
\end{equation}

In the situation where there are no spin flip events in the normal metal
($\lambda_N=\infty$) we find that we can write eq. \ref{Rspin} in an even more
simple form:

\begin{equation}
\Delta R =
\frac{2\alpha_F^2\lambda_F^2/\sigma_F^2}{(1-\alpha_F^2)^2SL/\sigma_N}\;.
\label{RspinSimple}
\end{equation}

The important point to notice is that Eq. \ref{RspinSimple} clearly shows that
even in the situation when there are no spin flip processes in the normal
metal, the spin signal $\Delta R$ is reduced with increasing $L$. The reason is
that the \emph{spin dependent} resistance ($\lambda_F/\sigma_{F}S$) of the
injecting and detecting ferromagnets remains constant for the two spin
channels, whereas the \emph{spin independent} resistance ($L/\sigma_{N}S$) of
the nonmagnetic metal in between the two ferromagnets increases linearly with
$L$. In both nonmagnetic metal regions II and V (Fig. \ref{diagram}) the spin
currents have to traverse a total resistance path over a length $\lambda_F+L/2$
and therefore the polarization of the current flowing through these regions
will decrease linearly with L and hence the spin signal $\Delta R$. Note that
in the regions V and VI no net current is flowing as the opposite flowing
spin-up and spin-down currents are equal in magnitude.

Using Eqs. \ref{currentup}, \ref{currentdown} and $I$ we can calculate the
current polarization \emph{at the interface} of the current injecting contact,
defined as
$P=~\frac{j_{\uparrow}^{int}-j_{\downarrow}^{int}}{j_{\uparrow}^{int}+j_{\downarrow}^{int}}$.
We obtain:

\begin{equation}
P=~\alpha_F\frac{Me^{L/2\lambda_N}+2cosh(L/2\lambda_N)}{2(M+1)
[Msinh(L/2\lambda_N)+cosh(L/2\lambda_N)]}\;. \label{polint1}
\end{equation}

In the limit that $L>>\lambda_N$ we obtain the polarization of the current at a
single F/N interface:\cite{son1}

\begin{equation}
P=~\frac{\alpha_F}{M+1}\;. \label{polint2}
\end{equation}

Again, Eq. \ref{polint2} shows a reduction of the polarization of the current
at the F/N interface, when the spin dependent resistance
($\lambda_F/\sigma_{F}S$) is much smaller that the spin independent resistance
($\lambda_N/\sigma_{N}S$) of the nonmagnetic metal. This situation becomes
progressively worse for a semiconductor as $\sigma_N$ is reduced by a factor of
$100$ or more and has become known as the "conductivity
mismatch".\cite{schmidt,filip1}

Finally we note that the spin signal $\Delta R^{Conv}$ can also be calculated
for a conventional measurement geometry, see Fig. \ref{sample}b, writing down
similar equations and boundary conditions as we have done for the non local
geometry (Eqs. $I$ to $VI$). We find:

\begin{equation}
\Delta R^{Conv} =~2\Delta R\;. \label{conventional}
\end{equation}

Equation \ref{conventional} shows that the magnitude of the spin valve signal
measured with a conventional geometry is increased with a factor two as
compared to the non local spin valve geometry (see also Ref. \onlinecite{fert2}
Eq. 45).

\section{Resistor model of multi-terminal spin valve
structures} \label{section4}

\begin{figure}[t]
\centerline{\psfig{figure=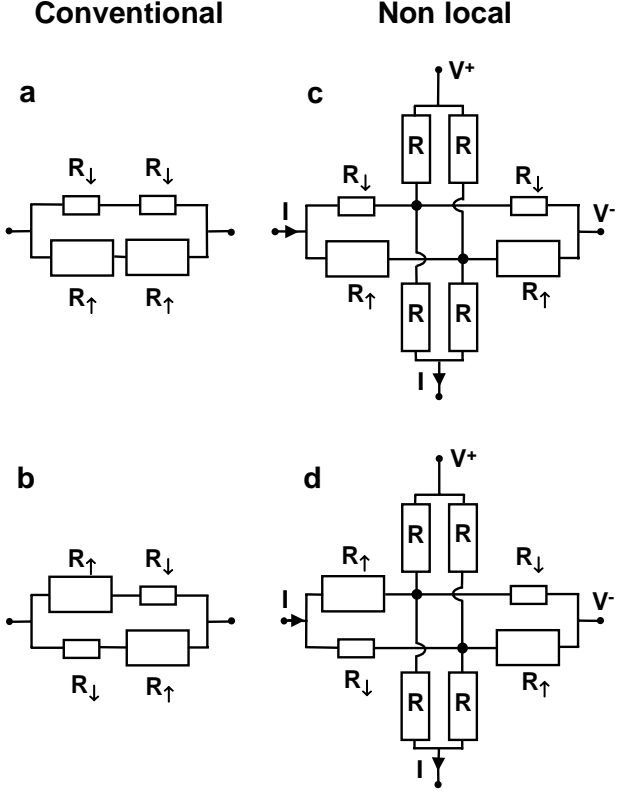,width=86mm}} \caption{The equivalent
resistor networks of the spin valve device. (a) The conventional spin valve
geometries in parallel and (b) in anti-parallel configuration. (c) The
non-local spin valve geometry in parallel and (d) in anti-parallel
configuration.} \label{model}
\end{figure}

More physical insight can be gained by considering an equivalent resistor
network of the spin valve device.\cite{lee1} In the linear transport regime,
where the measured voltages are linear functions of the applied currents, the
spin transport for the conventional and non local geometry can be represented
by a two terminal and four terminal resistor network respectively. This is
shown in Fig. \ref{model} for both parallel and anti-parallel configuration of
the ferromagnetic electrodes. The resistances $R_{\downarrow}$ and
$R_{\uparrow}$ represent the resistances of the spin up and spin down channels,
which consists of the different spin-up and spin-down resistance of the
ferromagnetic electrodes ($R^{F}_{\uparrow}, R^{F}_{\downarrow}$) and the spin
independent resistance $R^N$ of the nonmagnetic wire in between the
ferromagnetic electrodes. From resistor model calculations we obtain:

\begin{eqnarray}
R_{\uparrow} & = & R^{F}_{\uparrow}+R^N =
\frac{2\lambda_F}{w(1+\alpha_F)}R^{F}_\square + \frac{L}{w}R^{N}_\square
\label{Rup}\\
R_{\downarrow} & = & R^{F}_{\downarrow}+R^N =
\frac{2\lambda_F}{w(1-\alpha_F)}R^{F}_\square + \frac{L}{w}R^{N}_\square \;,
\label{Rdown}
\end{eqnarray}

where $R^{F}_\square=1/\sigma_Fh$ and $R^{N}_\square=1/\sigma_Nh$ are the
"square" resistances of the ferromagnet and non-magnetic metal thin films, $w$
and $h$ are the width and height of the nonmagnetic metal strip. The resistance
$R=(\lambda_N-L/2)2R^{N}_\square/w$ in Fig. \ref{model}c and Fig. \ref{model}d
represents the resistance for one spin channel in the side arms of the
nonmagnetic metal cross over a length $\lambda_N-L/2$, corresponding to the
regions IV and V of Fig. \ref{diagram}b.

Provided that $\lambda_N\gg L$ the spin dependent resistance $\Delta R^{Conv}$
between the parallel (Fig. \ref{model}a) and anti-parallel (Fig. \ref{model}b)
resistor networks for the conventional geometry can be calculated using eqs.
\ref{Rup} and \ref{Rdown}. We obtain the familiar expression:\cite{gijs1,bass1}

\begin{equation}
\Delta R^{Conv}
=~\frac{(R_{\downarrow}-R_{\uparrow})^2}{2(R_{\uparrow}+R_{\downarrow})}.
\label{ConvResistormodel}
\end{equation}

For the non local geometry and under the condition $\lambda_N\gg L$ the spin
dependent resistance $\Delta R$ between the parallel (Fig. \ref{model}c) and
anti-parallel (Fig. \ref{model}d) resistor network can also be calculated. We
obtain:

\begin{equation}
\Delta
R=~\frac{(R_{\downarrow}-R_{\uparrow})^2}{4(R_{\uparrow}+R_{\downarrow})}.
\label{NL}
\end{equation}

Equation \ref{NL} again shows that the spin signal measured in a non local
geometry is reduced by a factor 2 as compared to a  conventional measurement.
Provided that $R^{F}_{\uparrow}, R^{F}_{\downarrow} \ll R^{N}$ we can use Eqs.
\ref{Rup} and \ref{Rdown} to rewrite Eq. \ref{NL} into:

\begin{equation}
\Delta R =
\frac{2\alpha_F^2\lambda_F^2{R^{F}_\square}^2}{(1-\alpha_F^2)^2LwR^{N}_\square}\;.
\label{RspinSimpleResitorModel}
\end{equation}

Using $S=wh$ and replacing the square resistance by the conductivities Eq.
\ref{RspinSimpleResitorModel} reduces to Eq. \ref{RspinSimple}. A direct
relation can now be obtained between the experimentally measured quantities
$\Delta R$, $R^{N}_\square$, $R^{F}_\square$ and the relevant spin dependent
properties of the ferromagnet:

\begin{equation}
R_{\downarrow}-R_{\uparrow}=~\sqrt{8\Delta R R^{N}_\square \frac{L}{w}}=~
~{\frac{4 \alpha_F \lambda_F R^F_\square}{(1-\alpha_F^2)w}}.
\label{SpinResistance}
\end{equation}

\section{sample fabrication and measurement geometry} \label{geometry}

\begin{figure}[htp]
\centerline{\psfig{figure=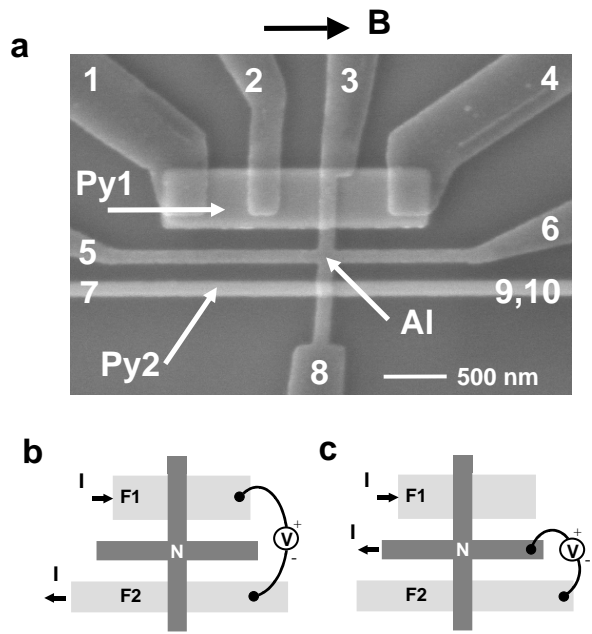,width=80mm}} \caption{a, Scanning
electron microscope (SEM) picture of the lateral mesoscopic spin valve device
with a ferromagnetic electrode spacing $L=500$ nm. The two horizontal strips
are the ferromagnetic electrodes F1 (Py1) and F2 (Py2). Their sizes are ($2
\times 0.5~\mu m^2$) and ($14 \mu m \times 0.15\mu m^2$) respectively. An
aluminum (Al) cross is placed in between the Py electrodes, which vertical arms
lay on top of the Py electrodes. A total of 10 contacts (not all visible) are
connected to the device. b, The conventional measurement geometry and (c) the
non local measurement geometry. The black arrow indicates the direction of the
applied magnetic field B in the measurements.} \label{sample}
\end{figure}

We use permalloy $Ni_{80}Fe_{20}$ (Py), cobalt (Co) and nickel (Ni) electrodes
to drive a spin polarized current into copper (Cu) or (Al) crossed strips.
Different aspect ratios of the rectangular ferromagnetic injector (F1) and
detector strips (F2) result in different switching fields of the magnetization
reversal process, allowing control over the relative magnetization
configuration of F1 and F2 (parallel/anti-parallel) by applying a magnetic
field parallel to the long axis of the ferromagnetic
electrodes.\cite{roukes1,nitta1} Two sets [F1,F2] of different sizes are used
in the experiments. One set has dimensions of $2 \times 0.8~\mu m^2$ (F1) and
$14\times 0.5~\mu m^2$ (F2), whereas the other set has dimensions of $2 \times
0.5~\mu m^2$ (F1) and $14\times 0.15~\mu m^2$ (F2). An example of a typical
device is shown in Fig. \ref{sample}.

The devices are fabricated in two steps by means of conventional e-beam
lithography with PMMA resist and lift-off technique. To avoid magnetic fringe
fields, the ferromagnetic electrodes are deposited first on a thermally
oxidized silicon substrate. The $40$ nm thick Py electrodes are sputter
deposited on a $2$ nm tantalum (Ta) adhesion layer. The base pressure of the
sputter system was $2\cdot 10^{-8}$ mbar vacuum, whereas the background Ar
pressure during sputtering was $1$ mbar. A small B-field of $~3$ mT along the
long axis of the Py electrodes was applied during growth in order to align the
magneto crystalline anisotropy with the magnetic shape anisotropy of Py
electrode. The conductivity of the Py film was determined to be
$\sigma_{Py}=~6.6\cdot 10^6 ~ \Omega ^{-1} m^{-1}$ and $\sigma_{Py}=~1.2\cdot
10^7 ~ \Omega ^{-1} m^{-1}$ at RT and 4.2 K respectively. The $40$ nm thick Co
($99.95~ \%~pure$) and $30$ nm thick Ni ($99.98~ \%~pure$) electrodes were
deposited by e-gun evaporation in a $1\cdot 10^{-6}$ mbar vacuum (base
pressure: $2\cdot10^{-7}$ mbar). The conductivities of the Co and Ni films were
determined to be $\sigma_{Co}=~4.2\cdot 10^6 ~ \Omega ^{-1} m^{-1}$ and
$\sigma_{Ni}=~7.6\cdot 10^6 ~ \Omega ^{-1} m^{-1}$ at RT, whereas at 4.2 K they
were $\sigma_{Co}=~7.3\cdot 10^6 ~ \Omega ^{-1} m^{-1}$ and
$\sigma_{Ni}=~1.6\cdot 10^7 ~ \Omega ^{-1} m^{-1}$. In the second fabrication
step, 50 nm thick crossed Cu ($99.99~ \%~pure$) or Al ($99.999~ \%~pure$)
strips were deposited by e-gun evaporation in a $1\cdot 10^{-8}$ mbar vacuum
(base pressure: $2\cdot10^{-9}$ mbar). Prior to the Cu or Al deposition, a few
nm of Py, Co or Ni material was removed from the ferromagnetic electrodes by
Kaufmann sputtering at $500$ Volt for $30$ seconds in a $2\cdot10^{-4}$ mbar Ar
pressure, thereby removing the oxide to ensure transparent contacts. The time
in between the Kaufmann sputtering and Cu or Al deposition was about 3 minutes.
The conductivities of the Cu and Al films were determined to be
$\sigma_{Cu}=~3.5\cdot 10^7 ~ \Omega ^{-1} m^{-1}$ and $\sigma_{Al}=~3.1\cdot
10^7 ~ \Omega ^{-1} m^{-1}$ at RT, whereas at 4.2 K they were
$\sigma_{Cu}=~7.1\cdot 10^7 ~ \Omega ^{-1} m^{-1}$ and $\sigma_{Al}=~8.0\cdot
10^7 ~ \Omega ^{-1} m^{-1}$.

Two different measurement geometries are used to measure the spin valve effect
in our device structure, the so called 'conventional' geometry
(Fig.\ref{sample}b) and 'non-local geometry (Fig.\ref{sample}c). In the
conventional measurement geometry the current is sent from contact $1$ to $7$
and the signal $R=V/I$ is measured between contacts $4$ and $9$, see Fig.
\ref{sample}a. The conventional geometry suffers from a relatively large
background resistance as compared to the spin valve resistance. Small parts of
the ferromagnetic electrodes underneath the vertical Cu or Al wires of the
cross are included in this background resistance, which can give rise to
AMR\cite{coehoorn1} contributions and Hall effects. In the non-local
measurement geometry the current is sent from contact $1$ to $5$ and the signal
$R=V/I$ is measured between contacts $6$ and $9$. This technique is similar to
the "potentiometric" method of Johnson used in Ref. \onlinecite{john1,john2}.
However, the separation of the current and voltage circuits, allow us to remove
the AMR contribution and Hall effecs of the ferromagnetic electrodes
completely: the (magneto) resistance of the current injecting contact (F1) is
not relevant because any voltage drop that develops across it will not
influence the current that is sent through it and similarly, no current flows
through the ferromagnetic voltage contact (F2), so its (magneto) resistance
does not effect the voltage measurement. Therefore the only region which could
possibly give rise to a magneto resistance signal is the nonmagnetic metal
cross, but explicit measurements confirm that this region does not produce any
magneto resistance. Moreover, this central region would be unable to produce a
Hall signal in a non local measurement with a magnetic field perpendicular to
the substrate plane as both voltage probes are attached longitudinal to the
direction the current.

\section{Spin accumulation in Py/Cu/Py spin valves} \label{pycu}

The measurements were performed by standard ac-lock-in-techniques, using
current magnitudes of $100~\mu A$ to $1$ mA. Typical spin valve signals of two
samples MSV1 and MSV2 (of the same batch) with a Py electrode spacing of
$L=~250$ nm are shown in the Figs. \ref{NonLocalCu},\ref{spinsmall} and
\ref{spinbig}. They are both measured in a non local measurement geometry and
conventional measurement geometry. Sample MSV1, data shown in Fig.
\ref{NonLocalCu} and \ref{spinsmall}, had a current injector Py1 electrode of
size $2 \times 0.5~\mu m^2$, whereas the detector electrode Py2 had a size of
$14 \times 0.15~\mu m^2$. Sample MSV2, data in shown Fig. \ref{spinbig}, had
wider Py electrodes of $2 \times 0.8~\mu m^2$ and $14 \times 0.5\mu m^2$. The
first set of (narrower) Py electrodes [Py1,Py2] had a more ideal switching
behavior and had three times larger switching fields as compared to the second
Set [Py1,Py2]. We note that a discussion of the magnetic behavior of the Py
electrodes and contacts has been given in Ref. \onlinecite{jedema2}. To control
the parallel/anti-parallel magnetization configuration of the Py electrodes a
magnetic field is applied parallel to the long (easy) axis of the Py
electrodes.

\subsection{Non local spin valve geometry}

\begin{figure}[htb] \centerline{\psfig{figure=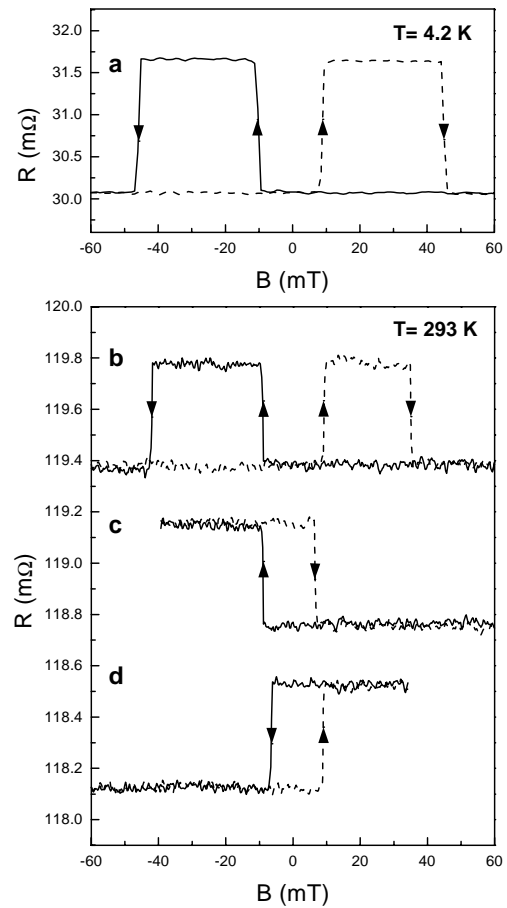,width=70mm}}
\caption{The spin valve effect at $T=4.2$ K (a) and RT (b) in the non-local
geometry for a Py/Cu/Py spin valve device (sample MSV1) with $250$ nm Py
electrode spacing. An increase in resistance is observed, when the
magnetization configuration is changed from parallel to anti-parallel. The
solid (dashed) lines correspond to the negative (positive) sweep direction.
(c),(d) illustrate the "memory effect". For clarity the (c) and (d) are off set
downwards. Note that the vertical scale of (a) is different from  (b),(c) and
(d). The sizes of the Py1 and Py2 electrodes are $2 \times 0.5~\mu m^2$ and $14
\mu m \times 0.15\mu m^2$} \label{NonLocalCu}
\end{figure}

Figure \ref{NonLocalCu}a and \ref{NonLocalCu}b show typical data in the non
local measurement geometry taken at 4.2 K and RT for sample MSV1 with a 250 nm
Py electrode spacing. Sweeping the magnetic field from negative to positive
field, an increase in the resistance is observed, when the magnetization of Py1
flips at $9$ mT, resulting in an anti-parallel magnetization configuration. The
rise in resistance is due to the spin accumulation or equivalently an excess
spin density present in the Cu metal. When the magnetization of Py2 flips at
$47$ mT ($T=4.2$ K) and $38$ mT (RT), the magnetizations are parallel again,
but now point in the opposite direction. The magnitude of the measured
background resistance, around $30~m \Omega$ at $T=4.2$ K and $120~m\Omega$ at
RT, depends on the geometrical shape of the Cu cross and is typically a
fraction of the Cu square resistance.

Figure \ref{NonLocalCu}c and \ref{NonLocalCu}d show the "memory effect". Coming
from high positive B field, the sweep direction of the B field is reversed
after Py1 has switched, but Py2 has not. At the moment of reversing the sweep
direction, the magnetic configuration of Py1 and Py2 is anti-parallel, and
accordingly a higher resistance is measured. When the B-field is swept back to
its original high positive value, the resistance remains at its increased level
until Py1 switches back at a positive field of 9 mT. At zero B field the
resistance can therefore have two distinct values, depending on the history of
the Py electrodes.

\subsection{Conventional spin valve geometry}

\begin{figure}[hbt] \centerline{\psfig{figure=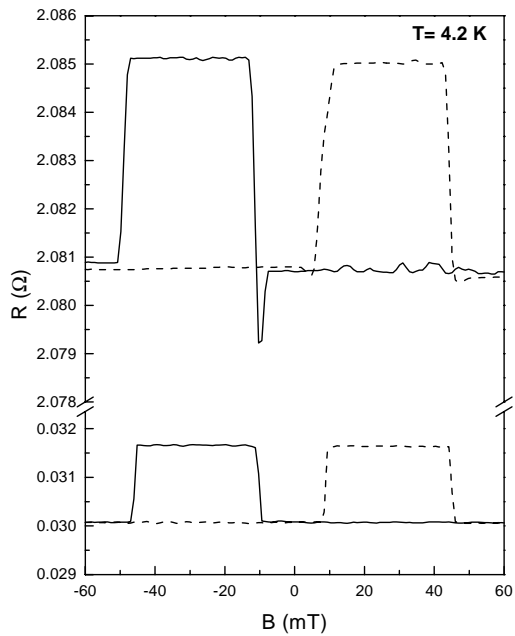,width=70mm}}
\caption{The spin valve effect of sample MSV1 in a conventional measurement
geometry (top curve) at $T= 4.2$ K and non-local measurement geometry (bottom
curve), with a Py electrode spacing $L= 250$ nm. The sizes of the Py electrodes
are $2 \times 0.5~\mu m^2$ (Py1) and $14 \times 0.15~\mu m^2$ (Py2). The
solid(dotted) curve corresponds with a negative (positive) sweep direction of
the B-field.} \label{spinsmall}
\end{figure}

The top curve in Fig. \ref{spinsmall} shows the magneto resistance behavior of
sample MSV1 in the conventional measurement geometry. A small AMR contribution
(dip in curve) of the Py1 electrode around $-9$ mT and a small Hall signal
caused by the Py2 electrode can be observed in the negative sweep direction.
Because a small part of the Py electrodes underneath the Cu wire is measured in
this geometry, (local) changes in the magnetization at the Py/Cu contact area
can produce an AMR or Hall signal.\cite{jedema2} In the positive sweep
direction a dip is no longer observed indicating that the magnetization
reversal of the Py1 electrode is not the same for the two sweep directions.
However, in the magnetic field range in between the two switching fields, we do
observe a resistance 'plateau' from $10$ mT up to a field of $45$ mT.

The magnitude of the spin valve effect measured in the conventional geometry is
about $4.1~m\Omega$ at $T=~4.2$. This is about $2.5$ times bigger than the
magnitude of the spin signal measured in a 'non-local geometry ($1.6~m\Omega$).
Note that the factor $2.5$ is deviating from the factor $2$ as predicted by the
eq. \ref{conventional}. This is attributed to deviations from our 1-dimensional
model, which can be expected for the samples with the shortest Py electrode
spacing $L=250$ nm, as the presence of the Cu side arms for these samples, see
Fig. \ref{sample}, are most felt.

\begin{figure}[htb] \centerline{\psfig{figure=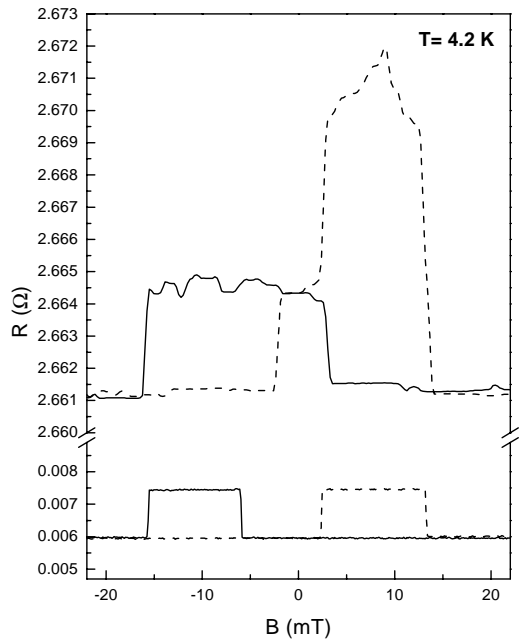,width=70mm}}
\caption{The spin valve effect of sample MSV2 in a conventional measurement
geometry (top curve) at $T= 4.2$ K and non-local measurement geometry (bottom
curve), with a Py electrode spacing $L= 250$ nm. The sizes of the Py electrodes
are $2 \times 0.8~\mu m^2$ (Py1) and $14 \times 0.5~\mu m^2$ (Py2). The
solid(dotted) curve corresponds with a negative (positive) sweep direction of
the B-field.} \label{spinbig}
\end{figure}

The top curve in Fig. \ref{spinbig} shows the magneto resistance behavior in
the conventional measurement geometry for the sample MSV2. Here a change of the
resistance is already observed before the field has reached zero in a positive
field sweep, whereas the negative field sweep is very asymmetrical compared to
the positive field sweep. This is attributed to the formation of a multi-domain
structure in the $2 \times 0.8~\mu m^2$ (Py1) electrode, causing a large AMR
($\approx10~m\Omega$) signal at the Py/Cu contact area of the Py1 electrode.

However, in a non local measurement geometry, the "contact" magneto resistance
contribution of the Py electrodes can be removed and a clear spin valve signal
is observed with a similar magnitude as sample MSV1. This is shown in the
bottom curve of Fig. \ref{spinbig}. Note also that the larger widths and aspect
ratio of the Py electrodes in sample MSV2 result in three times smaller
switching fields as compared to sample MSV1.

\subsection{Dependence on Py electrode spacing}\label{pycuspacing}

\begin{figure}[t]
\centerline{\psfig{figure=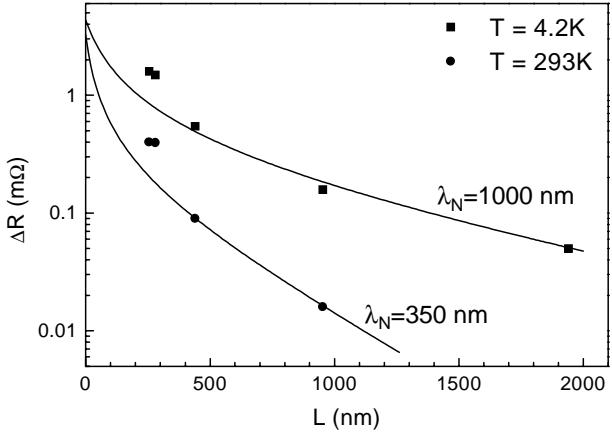,width=86mm}} \caption{Dependence of the
magnitude of the spin signal $\Delta R$ on the Py electrode distance $L$,
measured on Py/Cu/Py samples in the non local geometry. The solid squares
represent data taken at $T=~4.2$ K, the solid circles represent data taken at
RT. The solid lines represent the best fits based on equation \ref{Rspinfull}.}
\label{ldepentCu}
\end{figure}

A reduction of the magnitude of spin signal $\Delta R$ is observed with
increased electrode spacing $L$ , as shown in Fig. \ref{ldepentCu}. By fitting
the data to Eq. \ref{Rspinfull} we have obtained the spin relaxation length
$\lambda_{N}$ in the Cu wire. From the best fits we find a value of $1~\mu m$
at $T=~4.2$ K, and $350$ nm at RT. These values are compatible with those
reported in literature, where $450$ nm is obtained for Cu in GMR measurements
at $4.2$ K.\cite{yang1} However a detailed discussion on the obtained spin
relaxation lengths and corresponding spin relaxation times will be given in
Sec. \ref{spinrelaxation}.

In principle the fits of Fig. \ref{ldepentCu} also yield the spin polarization
$\alpha_F$ and the spin flip length $\lambda_{F}$ of the Py electrodes.
However, the values of $\alpha_F$ and $\lambda_F$ cannot be determined
separately, as in the relevant limit ($M~>>~1$) which applies to the Py/Cu/Py
experiments ($12<M<26$), the spin signal $\Delta R$ is proportional to the
product $\alpha_F\lambda_{F}$ as is shown by Eq. \ref{RspinSimple}. From the
fits we find that $\alpha_F\lambda_{F}=~1.2$ nm at 4.2 K and
$\alpha_F\lambda_{F}=~0.5$ nm at RT. Taking, from literature
\cite{dub1,steenwyk1,holody1}, a spin flip length in the Py electrode of
$\lambda_{F}=~5$ nm (at 4.2 K), a bulk current polarization of $\approx20~\%$
in the Py electrodes at $T=~4.2$ K is obtained: $\alpha_F=~0.2$. We note
however that the injected spin polarized current from the Py electrode is
partially shunted by the Cu wire lying on top of the Py electrode. When taken
into account we estimate that it could increase the value $\alpha_F\lambda_{F}$
by a factor $2$ to $3$.

It is also possible to calculate the polarization of the current at the Py/Cu
interface. For a sample with a Py electrode spacing of $L=250$ nm at $T=4.2$ K
and using Eq. \ref{polint1} we find: $P\simeq 0.02$, a factor $10$ lower than
the bulk polarization $\alpha_F$ of the Py electrodes. From the resistor model
we can see why the current polarization at the Py/Cu interface is reduced. For
this we need to calculate the magnitude of the spin dependent resistance
difference. Using Eq. \ref{SpinResistance} and $L=250$ nm, $\Delta
R=1.6~m\Omega$, $R^{N}_\square=0.3~\Omega$ and $w=100$ nm ( at $T=4.2$ K) we
find: $R_{\downarrow}-R_{\uparrow}\approx 100~m\Omega$. From the right hand
side term of Eq. \ref{SpinResistance} and using $R^{F}_\square=2~\Omega$ we can
check that this indeed corresponds with the value of
$\alpha_F\lambda_{F}\approx ~1.2$ nm, as was also obtained from the fit in Fig.
\ref{ldepentCu}. From Eqs. \ref{Rup}, \ref{Rdown} and using $\lambda_{F}=~5$
nm, $\alpha_F=~0.2$ (at 4.2 K) we obtain the spin up and spin down resistance
of the Py ferromagnet:

\begin{eqnarray}
R^{Py}_{\uparrow} & = & \frac{2\lambda_F}{w(1+\alpha_F)}R^{F}_\square \approx
160~m\Omega
\label{Rup2}\\
R^{Py}_{\downarrow} & = & \frac{2\lambda_F}{w(1-\alpha_F)}R^{F}_\square \approx
260~m\Omega \;. \label{Rdown2}
\end{eqnarray}

This shows that the total resistance experienced over a length
$\lambda_F+\lambda_N$ by the spin up and spin down currents is indeed dominated
by the spin \emph{independent} resistance $R^N+R=\lambda_N 2 R^{N}_\square /w
\simeq 6~\Omega$. Here we have used that $\lambda_N= 1~\mu m$ at $T=4.2$ K and
$w=100$ nm. This leads to a interface polarization of $P\approx
(R_{\downarrow}-R_{\uparrow})/(R^N+R) \approx 2\%$ at the Py/Cu interface.

All though the role of interface resistance for spin injection will be
described in the next section, we note here that the small difference
$R_{\downarrow}-R_{\uparrow}\approx 100~m\Omega$ responsible for a spin valve
signal of $\Delta R=1.6~m\Omega$ could possibly also result from an interface
resistance at the Py/Cu interface. Commonly reported resistivities of $5 \cdot
10^{-16}~\Omega m^2$ for the Py/Cu
interface\cite{dub1,steenwyk1,holody1,lee1,yang2} and a contact area of
$S=1\cdot10^{-14}~m^2$ (i.e. $R_{int}=50~m\Omega$) would yield a realistic
interface polarization of $\eta=0.5$ for the Py/Cu interface, using eq.
\ref{eta}. However, the specific details of the spin injection mechanism
(interface, bulk or a combination) do not alter the conclusion that the total
spin dependent resistance $R_{\downarrow}-R_{\uparrow}\approx 100~m\Omega$ is
dominated by the spin independent resistance of the Cu strip over a spin
relaxation length and hence leads to a considerable reduction of the spin valve
signal, as was pointed out above.

\subsection{Comparison with Johnson Spin Transistor}\label{sectionjohnson}

The magnitude of the spin signals in the Py/Cu/Py samples, when scaled to the
cross sections utilized in the Au thin film devices of Ref.
\onlinecite{john1,john2} (the "Johnson spin transistor"), are more than $10^4$
times smaller than obtained in that previous work. However, in that earlier
work it was necessary to invoke a spin polarization exceeding $100 \%$
to explain the results in terms of spin accumulation.\cite{john1,john2} This
contrasts our results, which yield a spin polarization P of the current
injected in the Cu wire at the Py/Cu interface of about $1-2 \%$.

In Refs. \onlinecite{john1,john2,john3} and \onlinecite{john4} Johnson
postulates that spin injection is mediated by interfacial transport, because
the interface resistances $R_{\uparrow}^{int}$, $R_{\downarrow}^{int}$ would
dominate the total resistance in both spin-up and spin-down channels:
$R_\uparrow^{int}
> R_\uparrow^F+R^N+R$ and $R_\downarrow^{int} > R_\downarrow^F+R^N+R$
respectively. In this limit spin injection would be characterized by the
interfacial spin injection parameter defined as:

\begin{equation}
\eta=\frac{R_{\downarrow}^{int}-R_{\uparrow}^{int}}{R_{\uparrow}^{int}+
R_{\downarrow}^{int}} \;,\label{eta}
\end{equation}

and Johnson derives the following expression for the spin accumulation
signal\cite{john2,john3}:

\begin{equation}
\Delta R = \frac{2\eta^2\lambda_N^2}{\sigma_NSL}\;.
 \label{Rspinjohnson}
\end{equation}

Applying Eq. \ref{Rspinjohnson} Johnson calculates an expected spin signal of
$\Delta R=1.9~\Omega$ for our device with the shortest Py electrode spacing
$L=250$ nm, using $S= 5\cdot 10^{-15}~m^2$, $\sigma_{Cu}=~7.1\cdot 10^7 ~
\Omega ^{-1} m^{-1}$, $\eta=0.4$ and $\lambda_N= 1.0~\mu m$.\cite{john3}

However a polarization of the current at the Py/Cu interface of $\eta=40~\%$
would require spin dependent interface resistances of
$R_{\uparrow}^{int}=16~\Omega$ and $R_{\downarrow}^{int}= 37~\Omega$, using
Eqs. \ref{eta}, \ref{NL} and replacing Eqs. \ref{Rup} and \ref{Rdown} by:

\begin{eqnarray}
R_{\uparrow} & = & R_{\uparrow}^{int}+R^{F}_{\uparrow}+R^N
\label{Rup3}\\
R_{\downarrow} & = & R_{\downarrow}^{int}+R^{F}_{\downarrow}+R^N  \;,
\label{Rdown3}
\end{eqnarray}
where the spin dependent interface resistances $R_{\uparrow}^{int}$ and
$R_{\downarrow}^{int}$ have simply been added up to bulk spin dependent
resistances $R^{F}_{\uparrow}$ and $R^{F}_{\downarrow}$ because the spin
polarization $\eta$ as well as the bulk spin polarization $\alpha_F$ are found
to be positive ($\alpha_F>0$ and $\eta>0$) for Py and
Cu.\cite{dub1,steenwyk1,holody1,vouille1} The values
$R_{\uparrow}^{int}=16~\Omega$ and $R_{\downarrow}^{int}= 37~\Omega$ yield a
total single interface resistance $R_{int} = 11~\Omega$ or equivalently, a
interface resistivity of $1 \cdot 10^{-13}~\Omega m^2$. This is more than a
$100$ times larger then the upper limit $0.1~\Omega$ or equivalently a contact
resistivity of ~$1 \cdot 10^{-15}~\Omega m^2$ that we are able to determine
from our experiment, see Figs. \ref{spinsmall} and \ref{spinbig}.

The above arguments also apply for the experiment of Refs.
\onlinecite{john1,john2} where a gold layer is sandwiched in between two Py
layers. There is no physical reason why there should exist an interface
resistivity larger than $1\cdot 10^{-13}~\Omega m^2$ between the Au and Py or
Co layers in the experiment of Ref. \onlinecite{john1}, which can explain an
interface current polarization of $\eta=0.4$ or more. Equation
\ref{Rspinjohnson} can therefore not be applied to the experiment of Ref.
\onlinecite{john1}, because it does not include the (fast) spin relaxation
reservoirs of the ferromagnetic injector and detector contacts, which dominate
the total spin relaxation in the case of transparent contacts, as was already
pointed out in Refs. \onlinecite{fert2,herschfield}.

In view of this, given the unexplained discrepancies ($\eta> 3)$ of the earlier
work in Ref. \onlinecite{john1,john2}, and the more consistent values obtained
in the recent work, it is our opinion that the results of Refs.
\onlinecite{john1,john2} cannot be reconciled with spin injection and spin
accumulation.

\section{Spin accumulation in Py/Al/Py spin valves}\label{pyal}

Here we will describe spin injection experiments using permalloy
$Ni_{80}Fe_{20}$ (Py) strips as ferromagnetic electrodes to drive a spin
polarized current via transparent contacts into aluminum (Al) crossed strips,
see Fig. \ref{sample}. Similar current polarizations and spin relaxation
lengths for Py and Al are obtained as in the previous section (Sec.
\ref{pycu}).

\subsection{Spin valve measurements}

\begin{figure}[hbt] \centerline{\psfig{figure=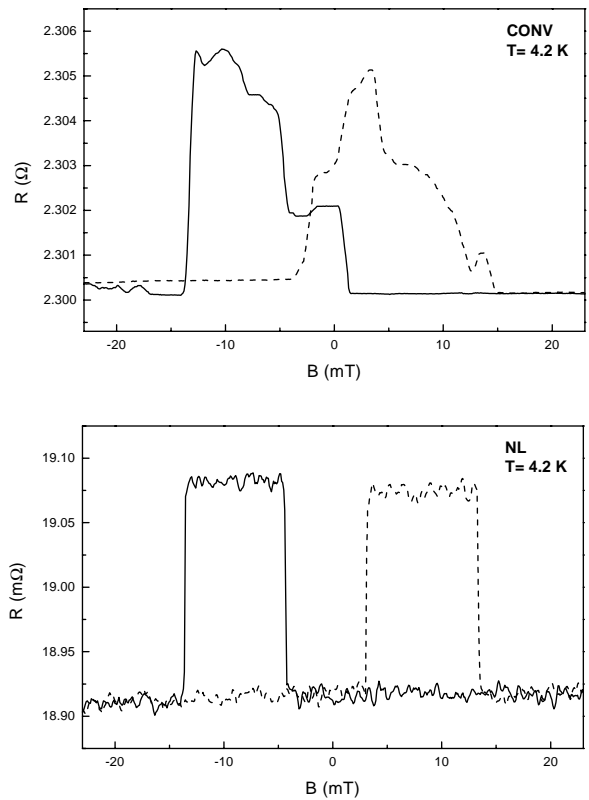,width=80mm}}
\caption{The spin valve effect of a Py/Al/Py sample using a conventional
measurement geometry (CONV, top curve) at $T= 4.2$ K and non-local measurement
geometry (NL, bottom curve), with a Py electrode spacing $L=250$ nm. The sizes
of the Py electrodes are $2 \times 0.8~\mu m^2$ (Py1) and $14 \times 0.5~\mu
m^2$ (Py2). The solid(dotted) curve corresponds with a negative (positive)
sweep direction of the B-field.} \label{spindataAl}
\end{figure}

Figure \ref{spindataAl} shows a typical spin valve signal of a Py/Al/Py sample
with a Py separation spacing of $L=250$ nm and Py electrodes of sizes $2 \times
0.8~\mu m^2$ and $14 \times 0.5~\mu m^2$.

The top curve in Fig. \ref{spindataAl} shows the magneto resistance behavior in
the conventional measurement geometry. Again the magneto resistance signals of
the Py contacts are dominating in this geometry, reaching a maximal amplitude
of about $6~m\Omega$. Note that the two resistance values at high positive and
negative field differ by a value of about $0.3~ m\Omega$, which is attributed
to a local hall effect caused by the $14 \times 0.5~\mu m^2$ Py electrode. The
bottom curves in Fig. \ref{spindataAl} show magnetic field sweeps in the non
local measurement geometry, which clearly shows a spin valve signal having
removed all the spurious contact magneto resistance effects. The magnitude of
the spin valve signal measured is $0.18~ m\Omega$ at $4.2$ K.

\subsection{Dependence on Py electrode spacing}

\begin{figure}[htb]
\centerline{\psfig{figure=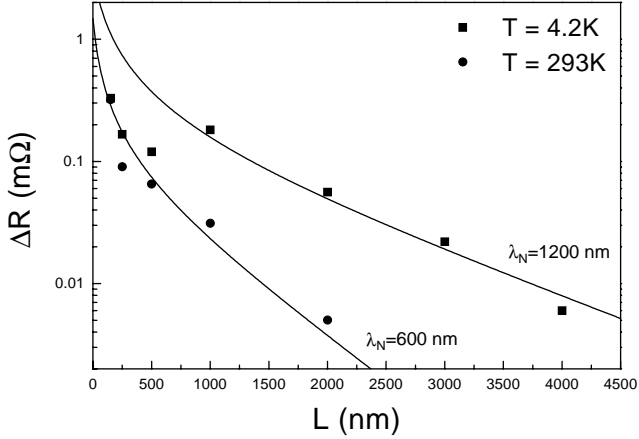,width=86mm}} \caption{Dependence of the
magnitude of the spin signal $\Delta R$ on the Py electrode distance $L$,
measured in the non local geometry for Py/Al/Py spin valves. The solid squares
represent data taken at $T=~4.2$ K, the solid circles represent data taken at
RT. The solid lines represent the best fits based on equation \ref{Rspinfull}.}
\label{ldepentAl}
\end{figure}

A reduction of the magnitude of spin signal $\Delta R$ of the Py/Al/Py samples
is observed with increased electrode spacing $L$, as shown in Fig.
\ref{ldepentAl}. However, for the $T=4.2$ K data this dependence is not
monotonic. The spin valve devices with small $L=250$ nm and $L=500$ nm show a
smaller spin valve signal than the device with $L=1~\mu m$. We note that all
the devices shown in Fig. \ref{ldepentAl} are from the same (processing) batch.
However, the granular structure of the Al film with a grain size in the order
of the width of the Al strip causes fluctuations in the resistance of the Al
strip in between the Py electrodes. The sample with $L=250$ nm and $L=500$ nm
indeed show a higher resistance than expected when measured in the conventional
geometry at $T=4.2$ K. This irregular behavior of the resistance due to grains
is not observed at Rt due to the additional presence of phonon scattering. From
the best fits to Eq. \ref{Rspinfull} we find a spin relaxation length
$\lambda_N$ in Al of $1.2~\mu m$ at $T=~4.2$ K and $600$ nm at RT. Note that
the spin flip lengths are about $2$ times larger than reported in Ref.
\onlinecite{jedema3}. The reason for this increase is the higher conductivity
of the Al in these samples, caused by a lower background pressure of $1 \cdot
10^{-8}$ mbar during evaporation as compared to a background pressure of $1
\cdot 10^{-6}$ used in Ref. \onlinecite{jedema3}.

The fits of Fig. \ref{ldepentAl} also yield the spin polarization $\alpha_F$
and the spin flip length $\lambda_{F}$ of the Py electrodes. We find
$\alpha_F\lambda_{F}=~1.2$ nm at 4.2 K and $\alpha_F\lambda_{F}=~0.5$ nm at RT,
in agreement with the Py/Cu/Py spin valve data of Sec. \ref{pycu}. Note that
for the Py/Al/Py spin valve also applies that $M~>>~1$ and thus the spin signal
$\Delta R$ is proportional to the product $\alpha_F\lambda_{F}$ ($25<M<32$).
Using Eq. \ref{polint1}, a polarization P for the Py/Al/Py sample with the
smallest Py electrode spacing of $L=~250$ nm at $T=4.2$ K is found to be only
$3 \%$: $P=0.03$.

\section{Spin injection using Co and Ni ferromagnetic electrodes}\label{CoNi}

From Eq. \ref{SpinResistance} it can be seen that the magnitude of the spin
dependent resistance $R_{\downarrow}-R_{\uparrow}$ is sensitive to the
properties $\alpha_F$, $\lambda_F$ and $\sigma_F$ of the ferromagnet. As
$R_{\downarrow}-R_{\uparrow}$ enters squared in the spin valve signal $\Delta
R$, see Eq. \ref{NL}, an increase of $\lambda_F$ with a factor $10$ would
increase $\Delta R$ with a factor $100$. We have therefore tried cobalt (Co)
and nickel (Ni) as ferromagnetic spin injectors and detectors to increase the
magnitude of the spin valve signal, as larger spin relaxation lengths can be
expected for these materials.\cite{ans1,bass1}

\subsection{Spin accumulation in Co/Cu/Co spin valves}

\begin{figure}[h]
\centerline{\psfig{figure=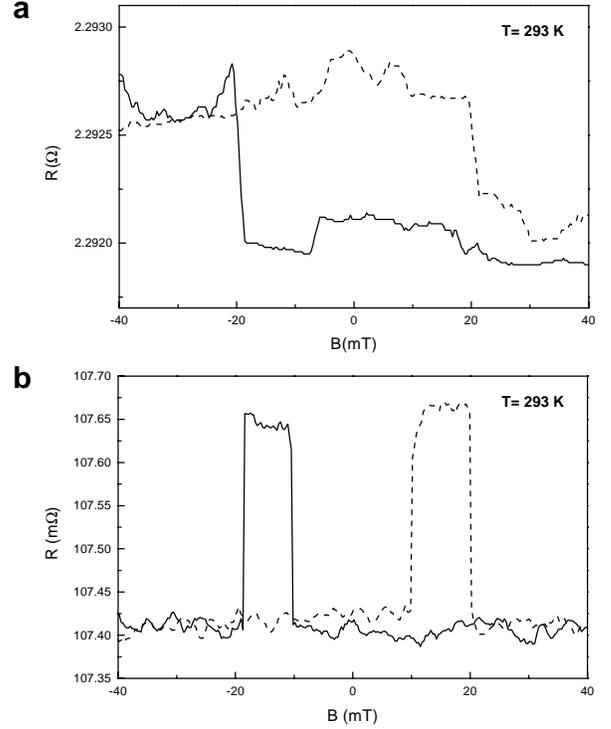,width=80mm}} \caption{a, "Contact"
magneto resistance trace of the Co2 electrode with size $14\times 0.5~\mu m^2$.
The Hall signal indicates an abrupt magnetization switching of the Co2
electrode. b, The spin valve effect at RT in a Co/Cu/Co device with a Co
electrode spacing $L= 250$ nm, using the non local measurement geometry. The
solid(dotted) curve corresponds with a negative (positive) sweep direction of
the B-field.} \label{cocudata}
\end{figure}

Figure \ref{cocudata}a shows a "contact" magneto resistance trace and magnetic
switching behavior at RT of a $14 \times 0.5~\mu m^2$ (Co2) electrode of a
Co/Cu/Co spin valve device with a Co electrode spacing of $250$ nm and Co
electrodes of sizes $2 \times 0.8~\mu m^2$ and $14 \times 0.5~\mu m^2$. The
"contact" magneto resistance is measured by sending current from contact $5$ to
$7$ and measuring the voltage between contacts $6$ and $9$ (see Fig.
\ref{sample}). Note that in this geometry the measured voltage is not sensitive
to a spin valve signal as only one Co electrode is used in the measurement
configuration. The magneto resistance traces of Fig. \ref{cocudata}a indicate a
clear switching of the magnetization at $20$ mT of the $14 \times 0.5~\mu m^2$
Co2 electrode and is attributed to a local Hall effect produced at the Co/Cu
contact area of the Co2 electrode.

Figure \ref{cocudata}b shows the spin valve effect at RT for a Co/Cu/Co spin
valve device. The magnitude of the spin dependent resistance $\Delta R=
0.25~m\Omega$ is slightly smaller than in the Py/Cu/Py spin valve device. At
$T=4.2$ K the signal increases to $\Delta R = 0.8~m\Omega$. Using Eq.
\ref{Rspinfull} and the values of $\sigma_N$, $\lambda_N$ for Cu and $\sigma_F$
for Co (see Sec. \ref{geometry}), we obtain $\alpha_F\lambda_F=0.3$ at RT and
$\alpha_F\lambda_F=0.7$ at $T=4.2$ K. These obtained values are much smaller
than reported for Co in GMR experiments, where $\alpha_F\approx 0.5$ and
$\lambda_F=10-60$ nm.\cite{dub1,lee2,piraux1,doudin1,ebels1} This discrepancy
will be discussed in Sec. \ref{sectionferro}.

\subsection{Spin accumulation in Ni/Cu/Ni spin valves}

\begin{figure}[hbt]
\centerline{\psfig{figure=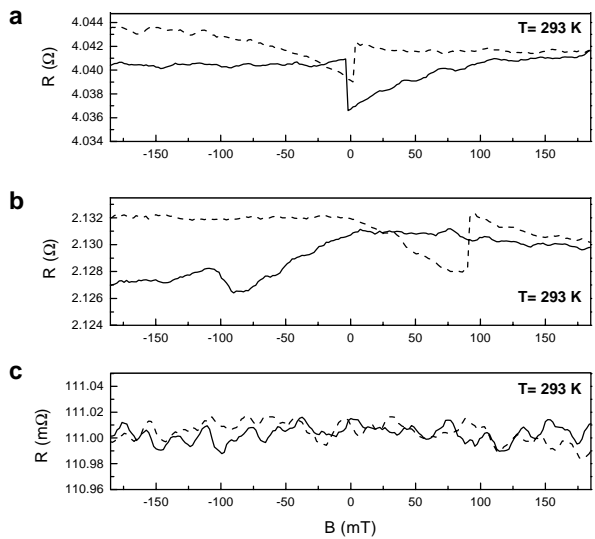,width=80mm}} \caption{(a) "Contact"
magneto resistance trace (see text) of the Ni1 electrode with size $2 \times
0.5~\mu m^2$. (b) "Contact" magneto resistance trace of the Ni2 electrode with
size $14 \times 0.15~\mu m^2$. (c)  The spin valve effect of a Ni/Cu/Ni device
at RT with a Ni electrode spacing of $L= 500$ nm, using a non local measurement
geometry. The solid(dotted) curve corresponds with a negative (positive) sweep
direction of the B-field.} \label{nicudata}
\end{figure}

In Fig. \ref{nicudata}a and b two "contact" magneto resistance traces of a Ni
electrode (Ni1) with size $2 \times 0.5~\mu m^2$ (top curve) and a Ni electrode
(Ni2) with size $14 \times 0.15~\mu m^2$ (middle curve) are shown of a Ni/Cu/Ni
spin valve device with a Ni electrode spacing of $500$ nm. For the Ni1 contact
current is send from contact $1$ to $5$ and the voltage is measured from
contact $4$ to $6$ (see Fig. \ref{sample}). For the Ni2 contact current is send
from contact $5$ to $7$ and the voltage is measured from contact $6$ to $9$. In
the magnetic field sweeps of Figs. \ref{nicudata}a and \ref{nicudata}b a large
range can be observed where the magnetization configuration of the Ni
electrodes are anti-parallel. However no spin valve signal could be detected
within experimental accuracy in the non local measurement geometry at RT as
well as at $T=4.2$ K, as is shown in Fig. \ref{nicudata}c (RT). An upper bound
on the spin valve signal is found to be $\Delta R <20~\mu\Omega$ at RT as well
as at $T=4.2K$. Using Eq. \ref{Rspinfull} and the values of $\sigma_N$,
$\lambda_N$ for Cu and $\sigma_F$ for Ni (see Sec. \ref{geometry}), we obtain
$\alpha_F\lambda_F<0.3$ at RT as well as at $T=4.2$ K. These obtained values
are also much smaller than reported for Ni in GMR experiments, where
$\alpha_F\approx 0.2$ and using the expected $\lambda_F(calc)=15$
nm.\cite{kubota1,vouille1} This discrepancy will be discussed in Sec.
\ref{sectionferro}.

We note that the magnetic field in the measurements of Fig. \ref{nicudata} is
applied perpendicular to the long axis of the Ni electrodes, showing a more
pronounced magnetic switching behavior than an applied magnetic field along the
long axis of the Ni electrodes.

\section{Spin relaxation times of conduction electrons in
metals}\label{spinrelaxation}

In this section we will analyze our obtained spin relaxation times $\tau_{sf}$
in Cu and Al from the spin injection experiments in Secs. \ref{pycu},\ref{pyal}
\ref{CoNi} and compare them with theory and previously reported values from
CPP-GMR\cite{pratt1}, CESR\cite{kittel1}, weak localization\cite{bergmann1} and
superconducting tunneling experiments\cite{tedrow1}. The obtained spin
polarization and spin relaxation lengths in Py, Co and Ni will be compared with
reported values from CPP-GMR experiments.

In CESR experiments the measured electron spin transverse relaxation time $T_2$
is proportional to the width of the absorption peak at the resonance frequency.
Yafet\cite{yafet1} showed that in metals $T_2$ is equal to the longitudinal
spin relaxation time or spin flip time $T_1$ ($T_1=\tau_{sf}$). In weak
localization and superconducting tunneling experiments the spin orbit
scattering time $\tau_{s.o.}$ is determined, with $\tau_{s.o.}$ being defined
similarly in both experiments.\cite{alexander1} Spin orbit interaction in weak
localization experiments is responsible for destructive interference when
electrons are scattered at (nonmagnetic) impurities\cite{bergmann1}, whereas in
the superconducting tunneling experiments it mixes up the spin-up and spin-down
quasi-particle density of states, when they are Zeeman-split by an applied
magnetic field.\cite{tedrow1,fulde1} We make the identification
$\tau_{s.o.}=\tau_{sf}$.

\subsection{Discussion of electron spin relaxation in nonmagnetic metals}

\begin{figure}[!hbt]
\centerline{\psfig{figure=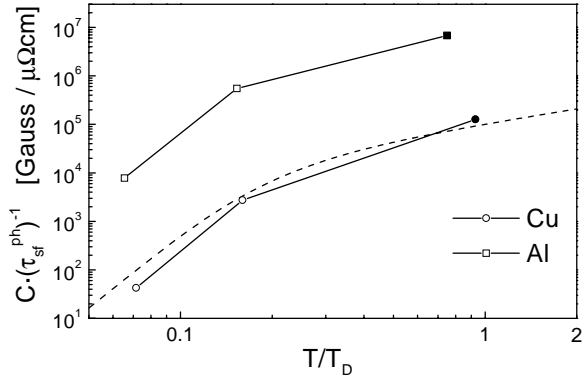,width=80mm}} \caption{The (revised)
Bloch-Gr$\mathrm{\ddot{u}}$neisen plot\cite{fabian1}. The quantity
$C\cdot(\tau_{sf}^{ph})^{-1}$ is plotted versus the reduced temperature $T/T_D$
on logarithmic scales. C represents a constant which links
$(\tau_{sf}^{ph})^{-1}$ to the (original) plotted width of a CESR resonance
peak, normalized by the spin orbit strength $(\lambda/\Delta E)^2$ and the
resistivity $\rho_D$ at $T=T_D$: $C=(\gamma(\lambda/\Delta E)^2\rho_D)^{-1}$.
Here $(\tau_{sf}^{ph})^{-1}$ is the phonon induced spin relaxation rate,
$\gamma$ is the Larmor frequency and $T_D$ is the Debye temperature. We used
$\rho_D=1.5\cdot10^{-8}$ and $T_D=315$ K for Cu and $\rho_D=3.3\cdot10^{-8}$
and $T_D=390$ K for Al.\cite{kittel1,bass2} The dashed line represents the
general Bloch-Gr$\mathrm{\ddot{u}}$neisen curve. The open squares represent Al
data taken from CESR and the JS spin injection experiment (Refs.
\onlinecite{lubzens1, silsbee1}). The open circles represent Cu data taken from
CESR experiments (Refs. \onlinecite{schultz1,beuneu2,monod2}). The solid square
(Al) and circle (Cu) are values from the spin injection experiments described
here and Refs. \onlinecite{jedema1,jedema3}.}\label{gruneisen}
\end{figure}

The fact that a spin can be flipped implies that there is some mechanism which
allows the electron spin to interact with its environment. In the absence of
magnetic impurities in the nonmagnetic metal, the dominant mechanism that
provides for this interaction is the spin-orbit interaction, as was argued by
Elliot and Yafet.\cite{elliot1,yafet1} When included in the band structure
calculation the result of the spin-orbit interaction is that the Bloch
eigenfunctions become linear combinations of spin-up and spin-down states,
mixing some spin-down character into the predominantly spin-up states and vice
versa.\cite{fabian1} Using a perturbative approach Elliot showed that a
relation can be obtained between the elastic scattering time $(\tau_e)$, the
spin relaxation time $(\tau_{sf})$ and the spin orbit interaction strength
defined as $(\lambda/\Delta E)^2$:

\begin{equation}
\frac{\tau_e}{\tau_{sf}}=a\propto(\frac{\lambda}{\Delta E})^2 \;,\label{elliot}
\end{equation}

where $\lambda$ is the atomic spin-orbit coupling constant for a specific
energy band and $\Delta E$ is the energy separation from the considered
(conduction) band to the nearest band which is coupled via the atomic spin
orbit interaction constant. Yafet has shown that Eq. \ref{elliot} is
temperature independent\cite{yafet1}. Therefore the temperature dependence of
$(\tau_{sf})^{-1}$ scales with the temperature behavior of the resistivity
being proportional to $(\tau_e^{-1})$. For many clean metals the temperature
behavior of the resistivity is dominated by the electron-phonon scattering and
can to a good approximation be described by the
Bloch-Gr$\mathrm{\ddot{u}}$neisen relation \cite{bass2}: $(\tau_{sf})^{-1}\sim
T^5$ at temperatures below the Debye temperature $T_D$ and
$(\tau_{sf})^{-1}\sim T$ above $T_D$. Using data from CESR experiments, Monod
and Beuneu \cite{beuneu1,monod1} showed that $(\tau_{sf})^{-1}$ follows the
Bloch-Gr$\mathrm{\ddot{u}}$neisen relation for monovalent alkali and noble
metals. In Fig. \ref{gruneisen} their results are replotted for Cu and Al,
using the revised scaling as applied by Fabian and Das Sarma.\cite{fabian1} In
addition we have plotted the obtained data points for Cu and Al at
$T/T_D\approx1$ from our spin injection experiment by calculating $\tau_{sf}$
from $\lambda_{sf}$ (see Sec. \ref{sectiontsf} below) and using the calculated
spin orbit strength parameters from Ref. \onlinecite{beuneu1}: $(\lambda/\Delta
E)^2=2.16\cdot10^{-2}$ for Cu and $(\lambda/\Delta E)^2=3\cdot10^{-5}$ for Al.

From Fig. \ref{gruneisen} it can be seen that for Cu the
Bloch-Gr$\mathrm{\ddot{u}}$neisen relation is well obeyed, \textit{including}
the newly added point deduced from our spin injection experiments at RT
($T/T_D=0.9$). For Al however the previously obtained data points as well as
the newly added point from the injection experiments at RT ($T/T_D=0.75$) are
deviating from the general curve, being about two orders of magnitude larger
than the calculated values based on Eq. \ref{elliot}. We note that we cannot
extract data for the Bloch-Gr$\mathrm{\ddot{u}}$neisen plot shown in Fig.
\ref{gruneisen} from our spin injection experiments at $T=4.2$ K, because the
impurity (surface) scattering rate is completely dominating the phonon
contribution at $T=4.2$ K.

Fabian and Das Sarma have resolved this discrepancy for Al by pointing out that
there can exist so called 'spin-hot-spots' at the Fermi surface of poly valent
metals(like Al). Performing an ab initio pseudopotential band structure
calculation of Al they showed that the spin flip contribution of these (small)
spin-hot-spot areas on the (large) Fermi surface dominate the total spin flip
scattering rate $\tau_{sf}$ , making it factor $100$ faster than expected from
the Elliot-Yafet relation.\cite{fabian2,fabian3,fabian4} A simplified reasoning
for the occurrence of these spin-hot-spots is that in poly valent metals the
Fermi surface can cross the first Brillioun zone making the energy separation
$\Delta E$ in Eq. \ref{elliot} between the (conduction) band to the spin orbit
coupled band much smaller at these (local) crossings and hence result in a
larger spin orbit strength $(\lambda/\Delta E)^2$. Our newly added data point
shows that the under estimation of the spin orbit strength also holds for Al at
RT ($T/T_D=0.75$), as can be seen in Fig.\ref{gruneisen}. However it is in
excellent agreement with the theoretical predicted value by
Fabian\cite{fabian4} as will show below.

\subsection{Quantative analysis of the spin relaxation time $\tau_{sf}$ in Cu and
Al}\label{sectiontsf}

Comparing the conductivities and spin relaxation lengths at RT and $T=4.2$ K we
can obtain the impurity and phonon scattering rate and their associated spin
relaxation rates. Therefore we can define an impurity spin relaxation ratio
$a^{imp}=\tau^{imp}/\tau^{imp}_{sf}$ and an inelastic (phonon) scattering ratio
$a^{ph}=\tau^{ph}/\tau^{ph}_{sf}$. Here $(\tau^{imp})^{-1}$ and
$(\tau^{ph})^{-1}$ are the impurity and phonon scattering rate and
$(\tau^{imp}_{sf})^{-1}$ and $(\tau^{ph}_{sf})^{-1}$ are the impurity and
phonon induced spin relaxation rate. From the measured conductivity at $T=4.2$
K and Eq. \ref{einstein} we can determine $\tau^{imp}$. Using the Mathiessen
rule $(\tau_e)^{-1}=(\tau^{imp})^{-1}+(\tau^{ph})^{-1}$ and the RT conductivity
we can determine $\tau^{ph}$. We note that the surface scattering in our
samples is dominating the impurity scattering, as the mean free paths of
$l_e\approx60$ nm for both Al and Cu at $T=4.2$ K are larger than their film
thicknesses ($50$ nm). In the calculation we use the free electron values
$N(E_F)_{Cu}= 1.8\cdot10^{28}$ states/eV/$m^3$ and $v_F(Cu)= 1.57\cdot10^6~m/s$
for Cu\cite{ashcroft} and we use $N(E_F)_{Al}= 2.4\cdot10^{28}$ states/eV/$m^3$
and $v_F(Al)= 1.55\cdot10^6~m/s$ for Al.\cite{constant}

\begin{table*}[thb] 
\begin{ruledtabular}
\centerline{\begin{tabular}{|l||c|c||c|c||c|}
\multicolumn{6}{c}{} \\
\multicolumn{6}{c}{Aluminum (Al)} \\
\multicolumn{6}{c}{} \\
\hline
                       & $\tau_{sf}^{imp}~[ps]$ & $a_{imp}$              & $\tau_{sf}^{ph}~[ps]$ & $a_{ph}$         &   Ref.         \\
\hline
      Theory           &         -         &    -                   &       $90$ \footnotemark[1]      &$1.2\cdot10^{-4}$ \footnotemark[1] &  \onlinecite{fabian4}    \\
\hline
   Spin Injection      &         $100$      &$0.6\cdot10^{-4}$       &      $85$ \footnotemark[1]      &$1.1\cdot10^{-4}$ \footnotemark[1]   &  \onlinecite{jedema3}   \\
\hline
   Spin Injection      &         $70$      &$3.7\cdot10^{-4}$       &      $124$ \footnotemark[1]      &$1.3\cdot10^{-4}$ \footnotemark[1]   &  This work   \\
\hline
   Spin Injection (JS)  &   $9000$ & $15\cdot10^{-4}$           & $4000$ \footnotemark[2] &$4.8\cdot10^{-4}$ \footnotemark[2] & \onlinecite{silsbee1}    \\
\hline
       CESR            &$3000-9000$   &$9.0\cdot10^{-4}$       &$1000-57000$ \footnotemark[3] &$2.6\cdot10^{-4}$ \footnotemark[3] & \onlinecite{beuneu2,fabian4,lubzens1}    \\
\hline
Anti-weak localization &        4-46       &$(0.2-1.2)\cdot10^{-4}$ &       -          &   -              &   \onlinecite{bergmann2,tinkham1}   \\
\hline
     Superconducting tunneling         &      8-160        &$(0.1-5)\cdot10^{-4}$   &    -             &  -               &   \onlinecite{tedrow1,meservey1,fulde2,monsma2}    \\
\hline
\multicolumn{6}{c}{} \\
\multicolumn{6}{c}{Copper (Cu)} \\
\multicolumn{6}{c}{} \\
\hline
   Spin Injection      &         $41$      &  $0.7\cdot10^{-3}$     &     $14$ \footnotemark[1]         &$2.0\cdot10^{-3}$ \footnotemark[1]  &  this work, \onlinecite{jedema1}   \\
\hline
       CESR            &$2000-9000$   &    $0.8\cdot10^{-3}$   &$2000-21000$ \footnotemark[4] &$1.1\cdot10^{-3}$  \footnotemark[4] & \onlinecite{schultz1,beuneu2}    \\
\hline
       GMR           &     $4$   &$19\cdot10^{-3}$       &   -   &  -     & \onlinecite{yang1}    \\
\hline
Anti-weak localization &        5          &   $1.3\cdot10^{-3}$    &       -          &   -              &   \onlinecite{bergmann2,tinkham1}   \\
\hline
Energy-level spectroscopy &      20-80       &   -     &       -          &   -              &   \onlinecite{ralph1}   \\
\hline
\multicolumn{6}{c}{} \\
\end{tabular}}
\caption{Comparison of spin relaxation times between different experiments.
$\tau_{sf}^{imp}~[ps]$ is the impurity induced spin relaxation time at low
temperatures $T\leq4.2$ K due to surface scattering, dislocations or grain
boundaries. $\tau_{sf}^{ph}~[ps]$ is the phonon induced spin relaxation time at
elevated temperatures due to inelastic phonon scattering. For definition of
$a_{imp}$ and $a_{ph}$ see text.}
\footnotetext[1]{For T=293 K} %
\footnotetext[2]{For T=45 K} %
\footnotetext[3]{For a temperature range T=[1..90] K} %
\footnotetext[4]{For a temperature range T=[1..60] K} \label{comparison}
\end{ruledtabular}
\end{table*}

The obtained parameters for Cu and Al ($\tau^{imp}_{sf}$,$\tau^{ph}_{sf}$,
$a^{imp}$, $a^{ph}$) are tabulated in tabel \ref{comparison}. From tabel
\ref{comparison} we see that $\tau_{sf}^{ph}$ and $a_{ph}$ for Al at RT are in
good agreement with the theoretical values as predicted in the bandstructure
calculation by Fabian and Das Sarma.\cite{fabian4} They are also in agreement
with the results obtained from CESR experiments and the earlier JS spin
injection experiments at temperatures below $90$ K. Note that the spin
relaxation times are $2$ orders of magnitude larger in those earlier
experiments due the use of extremely clean samples with electron mean free
paths of a few tens of micrometers. Also for Cu we see that $\tau_{sf}^{ph}$
and $a_{ph}$ at RT are in good agreement with the results obtained from CESR
experiments at temperatures below $60$ K.

The impurity scattering ratio $a^{imp}$ shows a much bigger spread in values
for both Al and Cu. We speculate that this due to the different origin of the
impurities in the samples used for the various measurement techniques. For the
CESR experiments the impurity scattering is caused by dislocations, whereas for
our experiment it is mainly due to surface scattering. In weak-localization and
superconducting tunnneling experiments the used films are even thinner then our
films (few nm), causing additional scattering from grain boundaries in addition
to the surface scattering.

We note that for thin films we use it is not possible to realize mean free
paths of the order of micrometers as they will always be limited by surface
scattering. However the sensitivity of the CESR technique does not allow
measurements of $\tau_{sf}^{ph}$ below typically $1$ ns, whereas the SQUID
detection technique used in Ref. \onlinecite{silsbee1} does not operate at RT.
Therefore spin injection into thin films is rather complementary to the CESR
techniques and the JS spin injection experiments in the determining
$\tau_{sf}^{ph}$ in the temperature range from liquid Helium to RT. Also, the
fact that about half of the momentum scattering processes at RT is due to
phonon scattering implies that the present obtained results on the spin
relaxation lengths in Al and Cu can be maximally improved by a factor of $2$ at
RT.

\subsection{Spin injection efficiency of Py, Co and Ni ferromagnets}
\label{sectionferro}

In addition to the spin-orbit spin scattering in metallic ferromagnets, as
described above for nonmagnetic metals, there is spin flip scattering by
magnons.\cite{fert4} In cobalt magnons are nearly absent at low temperatures
and only start to compete with the spin orbit spin flip scattering at
temperatures higher than $T=100$ K.\cite{piraux1} The spin flip scattering by
magnons has two effects. It will simply add to the spin orbit spin flip
scattering rate which reduces the spin relaxation length $\lambda_F$ of the
ferromagnet at higher temperatures. Secondly, it will lower the bulk current
polarization of the ferromagnet $\alpha_F$ by changing $\sigma_\uparrow$ and
$\sigma_\downarrow$ and in addition by giving rise to a "spin mixing rate"
which equalizes the spin-up and spin-down currents in the
ferromagnet.\cite{fert3,fert4} The presence of spin flip scattering by magnons
can therefore lower $\alpha_F$ as well as $\lambda_F$ at RT.

At low temperatures ($T<100 K$) and in absence of magnetic impurities an upper
estimate can be given for the expected spin relaxation length in Co and Ni due
to the spin orbit spin flip scattering only: $\lambda_F=\sqrt{D\tau_e/a}$,
where $a$ is taken from spin flip scattering cross-sections determined by CESR
experiments\cite{monod3,vouille1} and recently from magneto-optic
experiments\cite{koopmans1}: $a_{Fe}=1.1\cdot10^{-2}$, $a_{Ni}=1.5\cdot10^{-2}$
and $a_{Co}=4.2\cdot10^{-2}$. Using a free electron model, the spin relaxation
length $\lambda_{Py}$ for Py with $\sigma_{Py}=~8.1\cdot 10^6 ~ \Omega ^{-1}
m^{-1}$ and $\lambda_{Co}$ for Co with $\sigma_{Co}=~1.7\cdot 10^7 ~ \Omega
^{-1} m^{-1}$ have been estimated in this way in Ref. \onlinecite{dub1}:
$\lambda_{Py}(calc)\approx 9$ nm and $\lambda_{Co}(calc)\approx 36$ nm at
$T=4.2$ K.\cite{dub1} Note that $\lambda_F$ scales linearly with $\tau_e$ and
thus the conductivity of the ferromagnetic metal. In this respect the reported
value of $\lambda_{Co}=59$ nm in Ref.\onlinecite{piraux1} is quite remarkable,
because the conductivity of the Co metal ($\sigma_{co}=6.4\cdot 10^{6}$) used
in Ref.\onlinecite{piraux1} is about 3 times smaller then used to calculate
$\lambda_{Co}(calc)$ in Ref.\onlinecite{dub1} ($\sigma_{co}=1.74\cdot 10^{7}$),
which makes the expected $\lambda_{Co}(calc)=13$ nm in the experiment of
Ref.\onlinecite{piraux1}. For Ni we derive an estimate of $\lambda_F$ using a
free electron density of $5.4\cdot 10^{28}~m^{-3}$. With $\sigma_{Ni}=~1.6\cdot
10^7 ~ \Omega ^{-1} m^{-1}$ and $a_{Ni}=1.5\cdot10^{-2}$ we calculate:
$\lambda_{Ni}(calc)=15$ nm at $T=4.2$ K.

Because $M > 10$ for all our spin valve samples, we cannot separately determine
$\alpha_F$ and $\lambda_F$ from the magnitude of the spin valve signal $\Delta
R$. In table \ref{efficiency} we therefore give the "spin injection efficiency"
$\alpha_F\lambda_F$ together with reported values from GMR experiments. We note
that our thin film conductivities for Py, Co and Ni are within a factor $2$ of
the reported values in the GMR experiments.

\begin{table}[!htb]
\begin{ruledtabular}
\begin{tabular}{rlccccc} %

\multicolumn{1}{c}{}  &         \multicolumn{2}{c}{ $Ni_{80}Fe_{20}$ }   & \multicolumn{2}{c}{Co}  & \multicolumn{2}{c}{Ni}   \\
\cline{2-7}
    & \multicolumn{1}{|c|}{4.2 K}         &  \multicolumn{1}{c|}{RT}            &  \multicolumn{1}{c|}{4.2 K}      & \multicolumn{1}{c|}{RT}      &  \multicolumn{1}{c|}{4.2 K}    & \multicolumn{1}{c|}{RT}    \\
    &   \multicolumn{1}{|c|}{}    &    \multicolumn{1}{c|}{}        & \multicolumn{1}{c|}{}  &   \multicolumn{1}{c|}{} & \multicolumn{1}{c|}{}    &  \multicolumn{1}{c|}{}    \\
\cline{2-7}
$ \alpha_F\lambda_F$  & \multicolumn{1}{|c|}{1.2}   & \multicolumn{1}{c|}{0.5}    & \multicolumn{1}{c|}{0.7} &  \multicolumn{1}{c|}{0.3}  & \multicolumn{1}{c|}{$<$ 0.1}  &   \multicolumn{1}{c|}{$<$ 0.1}       \\
  MSV    &    \multicolumn{1}{|c|}{}     &    \multicolumn{1}{c|}{} &  \multicolumn{1}{c|}{} & \multicolumn{1}{c|}{} &   \multicolumn{1}{c|}{}  &    \multicolumn{1}{c|}{}  \\
\cline{2-7}
$ \alpha_F\lambda_F$    &  \multicolumn{1}{|c|}{3.6 - 4.0\footnotemark[1]}    & \multicolumn{1}{c|}{-}    & \multicolumn{1}{c|}{4.5 - 27.7\footnotemark[2]}  & \multicolumn{1}{c|}{8.1 - 15.5\footnotemark[2]}    & \multicolumn{1}{c|}{3\footnotemark[3]}  & \multicolumn{1}{c|}{-}   \\
  GMR   &  \multicolumn{1}{|c|}{}       &        \multicolumn{1}{c|}{}  &    \multicolumn{1}{c|}{}   &\multicolumn{1}{c|}{}&  \multicolumn{1}{c|}{}   &  \multicolumn{1}{c|}{}    \\

\end{tabular}
\caption{Spin injection efficiencies $\alpha_F\lambda_F$ in nm for three
different ferromagnetic metals. The data is deduced from the mesoscopic spin
valve (MSV) experiments with transparent contacts in a non local geometry using
Cu as nonmagnetic metal and
compared with results from GMR experiments.} %
\footnotetext[1]{From Refs.\onlinecite{dub1,steenwyk1,holody1}}%
\footnotetext[2]{From Refs.\onlinecite{ans1,bass1,lee2,piraux1,doudin1,ebels1}} %
\footnotetext[3]{From Refs.\onlinecite{kubota1,vouille1} ($\alpha_{Ni}=0.2$) and using $\lambda_{Ni}(calc)=15$ nm} %
\label{efficiency}
\end{ruledtabular}
\end{table}

Table \ref{efficiency} shows that our obtained spin injection efficiency of the
Py ferromagnet $\alpha_{Py}\lambda_{Py}$ is in quantative agreement with the
values reported in GMR experiments ($\alpha_{Py}=0.7$, $\lambda_{Py}=5~nm$),
taking into account that our obtained $\alpha_{Py}\lambda_{Py}$ represents a
minimal value due to a partially shunting of the injected current by the Cu
wire on top of the Py electrodes. The reduction of $\alpha_{Py}\lambda_{Py}$ at
RT beyond the ratio $1.8$ of the Py conductivies at $T=4.2$ K and RT could be
attributed to magnons lowering $\alpha_F$ at RT.

For the Co and the Ni ferromagnets we observe much smaller spin injection
efficiencies $\alpha_F\lambda_F$, being more than 1 order of magnitude smaller
than values of $\alpha_F\lambda_F$ obtained in GMR experiments. So the question
is, what is causing this rather large reduction of the the spin valve signal?

First we discuss the possible influences of an existing interface resistance at
the Co/Cu and Ni/Cu interfaces. From the resistance measured in a conventional
geometry we are able to determine an upper estimate of the (diffusive)
interface resistances. For the Co/Cu/Co spin valve of Fig. \ref{cocudata} we
find an upper limit for a single Co/Cu interface of $0.4~\Omega$, whereas for
the Ni/Cu/Ni spin valve of Fig. \ref{nicudata} we find for a single Ni/Cu
interface $0.6~\Omega$. We note that the associated interface resistivity
($\approx 5\cdot 10^{-15}~\Omega m^2$) values are about $5$ times larger than
calculated for Co/Cu (specular or diffusive)
interfaces\cite{gijs1,levy1,schep1,xia1} and also $5$ times larger than values
obtained from GMR experiments.\cite{yang2,lee2} In case these Co/Cu and Ni/Cu
interface resistances are spin dependent, the spin signal would be (largely)
increased as the sign of the bulk and interface spin asymmetries of Co, Ni and
Cu are found both to be positive\cite{vouille1,tsymbal1,schep1,xia1}
($\alpha_F>0$ and $\eta>0$). However this is clearly not observed. In the
opposite case of spin independent interface resistances, the interface
resistance for each spin channel ($\approx 1~\Omega$) will not reduce the
measured spin valve signal much as the spin independent interface resistance
just adds to the (larger) spin independent resistance of the Cu strip of about
$6~\Omega$ (see Sec. \ref{sectionjohnson} eqs. \ref{Rup3} and \ref{Rdown3}).
The spin signal can therefore only be significantly be reduced due to a
possible spin flip scattering mechanism at the interface, an effect which has
recently been studied in CPP-GMR spin valves.\cite{park1,marrows1} The physical
origin of this mechanism could be diverse, for instance: surface roughness
creating local magnetic fields due to the formation of random domains, a dead
magnetic layer of the first few nm of Co or Ni and the formation of
anti-ferromagnetic oxides CoO and NiO at the surface during the time in between
the Kaufmann sputtering and the Cu deposition. However as we do not have a
characterization of the interfacial structure we cannot analyze what could be
the most probable cause. We do note that the above mentioned mechanisms could
also apply for the Py/Cu interface, however the Py/Cu/Py spin valve data show
that their manifestation in these samples is apparently absent or less severe.

Secondly, a change in the bulk properties of the Co and Ni could explain the
small spin valve signals. All though in our opinion it is not likely that the
bulk spin relaxation length would be subdue to a substantial shortening, a
reduction of the polarization $\alpha_F$ in our Co and Ni ferromagnets might
occur. In CIP-GMR experiments\cite{egelhoff1} a strong decrease of more than an
order of magnitude in the GMR signal was reported upon changing the base
($H_2O$) pressure of the in the vacuum chamber from $10^{-8}$ to $10^{-5}$
mbar, just before deposition the Co and Cu layers. In our deposition chamber
the base pressure is only $10^{-7}$ mbar, whereas in the experiments e.g. on
Co/Ag multilayers\cite{lee2} the base pressure is of the system is $10^{-8}$
mbar. However, theoretical work\cite{tsymbal1} does predict
$\alpha_{Co}\approx0.6$ for Co with a conductivity close to our thin Co film
$\sigma_{Co}=~7.3\cdot 10^6 ~ \Omega ^{-1} m^{-1}$. We note however that the Co
layers in the Co/Cu multilayered nanowires\cite{piraux1} and the Co/Ag
multilayers\cite{lee2} have a hcp structure, whereas the calculations of Ref.
\onlinecite{tsymbal1} were done on fcc cobalt. Unfortunately we do not know the
crystallinity and/or the crystal orientation of our Co films.

\section{conclusions}

We have demonstrated spin injection and accumulation in metallic mesoscopic
spin valves with transparent contacts. We have shown that in a conventional
measurement geometry the magneto resistance effects of the injecting and
detecting contacts can be much larger than the spin valve effect, making it
impossible to observe the spin valve effect in a 'conventional' measurement
geometry. However, these contact effects can be used to monitor the
magnetization reversal process of the spin injecting and detecting contacts. In
a non-local measurement geometry we can completely isolate the spin valve
effect, as was reported earlier in Ref. \onlinecite{jedema1}. Using this
geometry we find spin relaxation lengths in Cu of around $1~\mu m$ at $T=~4.2$
K and $350$ nm at RT and spin relaxation lengths in Al of around $1.2~\mu m$ at
$T=~4.2$ K and $600$ nm at RT. The associated spin relaxation times in Al and
Cu are in good agreement with theory and values from experiments previously
reported in the literature. For the Py material we find spin relaxation lengths
and current polarizations in agreement with GMR experiments. However for Co we
obtain values of $\alpha_F\lambda_F$ which are up to a factor $40$ smaller than
their GMR counterpart. For Ni electrodes we are unable to resolve a spin valve
signal within the limits of our experimental accuracy, corresponding with
$\alpha_F\lambda_F$ at least a factor $10$ lower than expected. Finally, we
believe that the use of tunnel barriers should make it possible to increase the
polarization of the injected current in nonmagnetic metals, as we recently have
shown.\cite{jedema3,jedema4} This should make it possible to increase the spin
signals  to about $1~\Omega$ in metals.

The authors wish to thank H. Boeve, J. Das and J. de Boeck at IMEC (Belgium)
for support in sample fabrication and the Stichting Fundamenteel Onderzoek der
Materie for financial support and J. Fabian for making his data on spin
relaxation times available to us.


\begin{references}
\bibitem{notation1} In our definition $\tau_{sf}$ is equal to the spin lattice
relaxation time $T_{1}$ as used in the Bloch equations.\cite{silsbee2} In the
Valet-Fert theory $(\tau_{sf})^{VF}$ is different. The difference originates a
different expression for $(\lambda_{sf}^2)^{-1}$. In the V-F theory
[\onlinecite{valet1}]
$\lambda_{sf}^2$=$(1/D_\uparrow+1/D\downarrow)^{-1}\tau_{sf}^{VF}$, whereas we
derive $\lambda_{sf}^2$=$([N_\downarrow/D_\uparrow(N_\uparrow+N_\downarrow)
+N_\uparrow/D_\downarrow(N_\uparrow+N_\downarrow)]^{-1}\tau_{sf}$. In
nonmagnetic metals the definition of [\onlinecite{valet1}] corresponds
therefore to twice the value of our $\tau_{sf}$:
$(\tau_{sf})^{VF}=2\tau_{sf}=2T_{1}$.
\end{references}
\end{document}